\documentclass[a4paper,11pt]{article}

\usepackage{jheppub}
\usepackage[T1]{fontenc}
\usepackage{graphicx}
\usepackage{amsmath}
\usepackage{amsmath}
\usepackage{xspace}
\usepackage{subfig}
\usepackage{color}
\usepackage[utf8]{inputenc}
\usepackage{commath}
 \usepackage{cancel}
\usepackage{slashed}
\usepackage{fancyvrb}
\usepackage{multirow}
\def\hc{{\rm h.c.}}

\title{Testing Leptogenesis and Seesaw in the $B-L$ Model using Long-lived Particle Searches}

\author[a]{Wei Liu,}
\author[b]{Frank F. Deppisch}
\author[a]{Zixiang Chen}

\affiliation[a]{Department of Applied Physics and MIIT Key Laboratory of Semiconductor Microstructure and Quantum Sensing, Nanjing University of Science and Technology, Nanjing 210094, China}
\affiliation[b]{University College London, Gower Street, London WC1E 6BT, UK}

\emailAdd{wei.liu@njust.edu.cn}
\emailAdd{f.deppisch@ucl.ac.uk}

\abstract{
We discuss the potential of using long-lived particle~(LLP) searches for right-handed neutrinos~(RHNs) to test resonant leptogenesis and the seesaw mechanism. This is challenging if only RHNs are added to the Standard Model~(SM), as naturally the active-sterile mixing strengths $|V_{\ell N}|^2$ are small, for 1~GeV~$\lesssim M_N \lesssim 1000$~GeV. Instead, we consider the minimal $B-L$ gauge model, where a $Z^\prime$ gauge boson couples to fermions including the RHNs. During leptogenesis, this gauge coupling introduces scattering processes that washout the $B-L$ asymmetry. At colliders, it can lead to abundant production of RHNs which allows probing the associated seesaw mechanism using LLP searches. We find that LLP searches at the FCC-hh can test leptogenesis and the seesaw mechanism simultaneously, and probe the active-sterile mixing at or below the seesaw floor.
}

\begin{document}
\maketitle
\flushbottom
\setcounter{footnote}{0}

\section{Introduction}
\label{sec:intro}
The baryon asymmetry of the Universe~(BAU) is arguably the most obvious evidence for physics beyond the Standard Model~(SM). As a popular solution, leptogenesis connects the BAU to light neutrino masses, which is another feature unexplained in the SM~\cite{Fukugita:1986hr, Luty:1992un, Davidson:2008bu}. Both phenomena suggest the existence of right-handed neutrinos~(RHNs) \cite{Minkowski:1977sc}. If the relevant neutrino mass terms contained $CP$ phases, the decays of the RHNs introduce a lepton asymmetry, which converts to the BAU via electroweak~(EW) sphaleron processes. In the standard leptogenesis mechanism, the magnitude of the $CP$ asymmetry depends on the masses of the lightest RHNs. In order to explain the observed BAU, the masses of the RHNs are required to be $M_N \gtrsim 10^9$~GeV~\cite{Davidson:2002qv}. Hence, it is not possible for such high-scale leptogenesis scenarios to be tested at colliders. However, if at least two RHNs are degenerate, the $CP$ asymmetry can be resonantly enhanced \cite{Flanz:1996fb, Pilaftsis:1997jf, Pilaftsis:2003gt, Iso:2010mv, Dev:2017wwc}. Therefore, in such resonant leptogenesis mechanisms, the observed BAU can be explained for $M_N$ as low as $M_N \gtrsim \mathcal{O}(100)$~GeV~\footnote{Successful leptogenesis can occur for even lower RHN masses via oscillations~\cite{Akhmedov:1998qx, Asaka:2005pn}.}.

In the type-I seesaw mechanism, the masses of the RHNs are connected to the light neutrino masses $m_\nu$ via the active-sterile mixing $V_{\ell N}$, $m_\nu \approx |V_{\ell N}|^2 M_N$. Since the observed neutrino masses are small, $\Sigma_i m_{\nu_i} \approx$ 0.06 eV for the normal ordering and $m_{\nu_1} = 0$, $V_{\ell N}^2 \lesssim 10^{-12}$ for $M_N \sim \mathcal{O}(100)$~GeV. We refer to this regime $m_\nu = |V_{\ell N}|^2 M_N$ as the canonical 'seesaw floor'. Due to their phenomenological importance, analyses and searches for RHNs have been carried out in numerous experimental and theoretical contexts \cite{Deppisch:2023sga, Li:2023dbs, Barducci:2023hzo, Liu:2023gpt, Liu:2023nxi, Zhang:2023nxy, Mandal:2023mck, Duarte:2023tdw, Arun:2022ecj, Delgado:2022fea, Liu:2021akf, Abdullahi:2022jlv, ThomasArun:2021rwf, Balaji:2020oig,Barducci:2022gdv, Ding:2019tqq,Shen:2022ffi, Beltran:2022ast,Zhou:2021ylt, Abada:2018sfh,Fernandez-Martinez:2022gsu, Abada:2022wvh, Arganda:2015ija, Bai:2022lbv, Das:2017nvm, Das:2015toa, Das:2016hof, Izmaylov:2017lkv, Batell:2016zod, Bhattacherjee:2021rml, Accomando:2017qcs, Das:2019fee, Cheung:2021utb,Chiang:2019ajm, FileviezPerez:2020cgn, Das:2018tbd, Han:2021pun, Mason:2019okp, Accomando:2016rpc, Gao:2019tio, Gago:2015vma, Jones-Perez:2019plk}. The allowed parameter space of sterile RHNs, as constrained by experimental searches, is, for example, summarised in  Refs.~\cite{Bolton:2019pcu, Abdullahi:2022jlv}, in terms of ($M_N$, $|V_{\ell N}|^2$). What makes the sterile RHN scenarios interesting is that the probed parameter space, considering the sensitivity of future searches can be mapped to the parameter space for successful resonant leptogenesis, \cite{Chun:2017spz, Antusch:2017pkq, Klaric:2020phc, Klaric:2021cpi, Drewes:2021nqr}. However, current and future searches are unlikely to probe the canonical seesaw floor. In the minimal scenario, the production of the RHNs is directly related to the magnitude of the active-sterile mixing, which is not sufficient to generate experimental signatures except for very specific cases~\cite{Deppisch:2015qwa}. 

Extending the SM with an additional gauge symmetry, the minimal $B-L$ model can explain the Majorana masses of the RHNs, as required for the type-I seesaw mechanism, via the spontaneous symmetry breaking of the $B-L$ gauge symmetry~\cite{Davidson:1978pm, Marshak:1979fm, Mohapatra:1980qe, Davidson:1987mh}. In this model, additional RHN production channels are present via decays of the $B-L$ scalar as well as the new gauge boson $Z^\prime$. Their production rate is independent of the active-sterile mixing, whereas the RHN decays still depend on it. If their decay can be detected, e.g., in LLP searches, we can probe the seesaw floor in a much broader range~\cite{Deppisch:2018eth, Deppisch:2019kvs, Liu:2022kid, Liu:2022ugx, Deppisch:2013cya, Batell:2016zod, Bhattacherjee:2021rml, Accomando:2017qcs, Das:2019fee, Cheung:2021utb, Chiang:2019ajm, FileviezPerez:2020cgn, Amrith:2018yfb, Das:2018tbd, Han:2021pun, Das:2017nvm, Pilaftsis:1991ug, Graesser:2007yj, Maiezza:2015lza, Nemevsek:2016enw, Mason:2019okp, Accomando:2016rpc, Gao:2019tio, Gago:2015vma, Jones-Perez:2019plk}. An experimental search for the pair-production of RHNs in $Z^\prime$ has also been performed at the CMS experiment~\cite{CMS:2023ooo}. For small active-sterile mixing, the RHNs are likely to be long-lived, therefore becoming one of the most promising long-lived particle~(LLP) candidates. Searches for LLPs have drawn a lot of attention recently, cf. Ref.~\cite{Alimena:2019zri} for a review. Targeting on searching such LLPs, a series of proposed detectors have been put forward. Among them, 
FASER~\cite{FASER:2018eoc} are already installed for Run 3 of the LHC. In the future, the LHC might be replaced by a future 100 TeV circular collider~(FCC-hh), with an expected improved reach~\cite{Liu:2022kid}.

It is thus natural to consider whether we can use LLP searches to probe and test leptogenesis and reach the canonical seesaw floor at colliders in the $B-L$ model. Nevertheless, once processes such as $Z^\prime \leftrightarrow NN$ are introduced in the $B-L$ model, they also lead to an additional lepton asymmetry washout~\cite{Blanchet:2009bu, Okada:2012fs, Deppisch:2013jxa, Heeck:2016oda, Dev:2017xry, BhupalDev:2014hro, BhupalDev:2015khe, BhupalDev:2019ljp, Liu:2021akf}. Previous works focused on testing the parameter space of the $B-L$ via successful leptogenesis. Nevertheless, whether leptogenesis and the canonical seesaw floor can be probed simultaneously has not been discussed extensively. We demonstrate that both can be probed by LLP searches at the FCC-hh. This is not trivial; while LLP RHNs are produced more abundantly that can be detected at or below the seesaw floor, a corresponding larger washout may jeopardize leptogenesis. In this work, we investigate whether LLP searches can probe the canonical seesaw floor and leptogenesis simultaneously. Based on the minimal $B-L$ model, the additional scattering processes due to $Z^\prime \leftrightarrow NN$  are considered. The maximal $CP$ asymmetry generated by resonant leptogenesis is also calculated, and shown that a detectable magnitude, $\epsilon_{\text{max}} > 0.1$ is found near the seesaw floor as well. We combine the parameter space in ($M_N$, $|V_{\ell N}|^2$) of successful leptogeneis, canonical seesaw mechanism and the sensitivity from LLP searches for $pp \to Z^\prime \to NN$ at the FCC-hh.

This paper is organized as follows. In Section~\ref{sec:lep}, we briefly introduce the $B-L$ model, and the resonant leptogenesis mechanism in its context. The parameter space of successful leptogenesis is compared with LLP searches at the FCC-hh in Section~\ref{sec:llp}. Finally, we conclude in Section~\ref{sec:con}.

\section{Theoretical Framework}
\label{sec:lep}

\subsection{The Minimal $B-L$ Gauge Model}
Before discussing leptogenesis, we briefly review the relevant details of the $B-L$ model.
In addition to the SM particle content, the minimal $B-L$ model contains new particles, namely the $B-L$ gauge boson $Z^\prime$, the $B-L$ scalar $\Phi$, and three RHNs $\nu_R^i$. The relevant Lagrangian reads 
\begin{equation}
\label{LB-L}
    \begin{split}
        \mathcal{L}_{B-L}=&~\sum_{i}\bar\nu_R^ii\slashed{D}\nu_R^i-\frac12\sum_{i,j}\left(\lambda_N^{ij}\bar\nu_R^{i,c}\Phi\nu_R^j+\hc\right)
    -\sum_{i,j}\left(\lambda_D^{ij}\bar\ell_L^i\tilde H\nu_R^j+\hc\right)\\
    &~+D_\mu\Phi^\dagger D^\mu\Phi-V(H, \Phi)-\frac14Z'_{\mu\nu}Z'^{\mu\nu}.
\end{split}
\end{equation}
Here, the family indices are represented by $i, j$. $D_\mu = D_{\mu, \text{SM}} - i g_{B-L} Q_{B-L} Z_\mu'$ is the covariant derivative incorporating the $B-L$ contribution to the SM part, i.e., $D_{\mu, \text{SM}}$, where $g_{B-L}$ and $Q_{B-L}$ are the gauge coupling and $B-L$ charge, respectively. $Q_{B-L}$ equals the baryon number minus the lepton number for all SM particles whereas for the exotic particles it is defined by $Q_{B-L}(\nu_R^i) =-1$, $Q_{B-L}(\Phi) = 2$. The scalar potential $V(H, \Phi)$ is constructed from all gauge invariant terms involving the SM Higgs doublet and $\Phi$.

After spontaneous symmetry breaking of the $B-L$ gauge, the $Z^\prime$ and RHNs obtain masses,
\begin{equation}
    M_{Z'}=2g_{B-L}v_\phi,\quad M_{N_i} = \lambda_{N_i}\frac{v_\phi}{\sqrt2},
\end{equation}
where $v_\phi$ is the vacuum expectation value~(vev) of $\Phi$, and $\lambda_{N_i}$ are the Majorana Yukawa couplings after diagonalizing the mass eigenstates.

\begin{figure}[t!]
    \centering
    \includegraphics[width=0.49\textwidth]{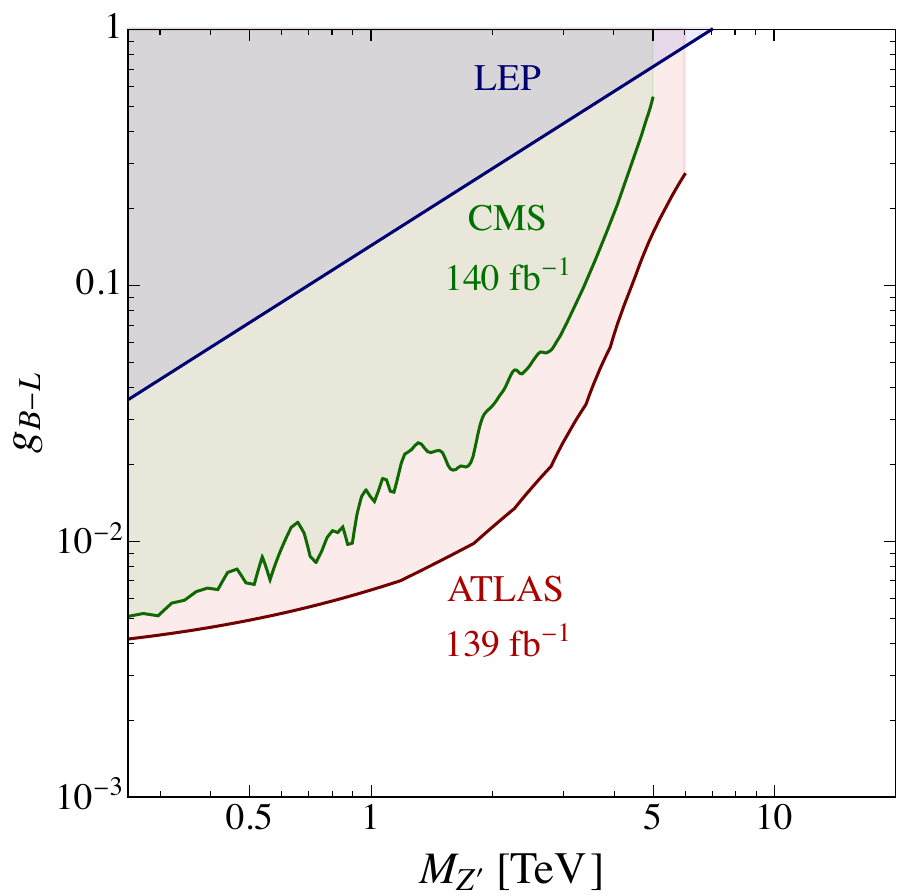}
    \caption{Current limits on the $Z^\prime$ gauge boson mass $M_{Z^\prime}$ and gauge coupling $g_{B-L}$ in the $B-L$ model~\cite{Liu:2022kid,Chiang:2019ajm}, recast from CMS/ATLAS searches for high-mass dilepton resonances, using LHC Run 2 data~\cite{ATLAS:2019erb, CMS:2021ctt}, and EW precision test limits mainly from LEP~\cite{Cacciapaglia:2006pk, ALEPH:2006bhb}.}
\label{fig:limit}
\end{figure}
The current limits on $M_{Z^\prime}$ and $g_{B-L}$ are summarised in Fig.~\ref{fig:limit}, for heavy $Z^\prime$ with $M_{Z^\prime} > 200$~GeV. The limits mainly derive from CMS/ATLAS searches for high-mass di-lepton resonances, using LHC Run 2 data~\cite{ATLAS:2019erb, CMS:2021ctt}, and EW precision test limits from LEP~\cite{Cacciapaglia:2006pk, ALEPH:2006bhb}. It can be seen that $M_{Z^\prime} \gtrsim 6$~TeV is weakly constrained by current experiments. To compare with the results in Ref.~\cite{Liu:2022kid}, we fix $M_{Z^\prime} = 5$~TeV, so the most stringent limits come from the ATLAS di-lepton searches, with $g_{B-L} \lesssim 0.16$.

The breaking of the $B-L$ gauge symmetry and the generating of heavy Majorana masses for the RHNs gives rise to the seesaw mechanism. Working in a basis where the charged lepton and RHN mass matrices are diagonal, the flavour structure is controlled by the Dirac Yukawa matrix $\lambda_D$. Being a general, complex $3\times 3$ matrix, it contains 18 parameters in total. To satisfy the neutrino oscillation data, we express $\lambda_D$ using the Casas-Ibarra parametrization~\cite{Casas:2001sr},
\begin{equation}
    \lambda_D = \frac{1}{v_{\text{EW}}} U \sqrt{\hat{m_\nu}} R^T \sqrt{\hat{M_N}},
\end{equation}
where $v_\text{EW} = 174$~GeV is the EW vev, $\hat{m_\nu} / \hat{M_N}$ is a diagonal matrix containing the light/heavy neutrino masses, and $U$ is the PMNS matrix, for which we adopt the PDG convention~\cite{ParticleDataGroup:2016lqr},
\begin{equation}
\begin{aligned}
    U =&\begin{pmatrix}
        1 & 0 & 0 \\
        0 & c_{23} & s_{23}  \\
        0 &- s_{23} & c_{23} 
    \end{pmatrix}
    \begin{pmatrix}
        c_{13} & 0 & s_{13}e^{-i\delta} \\
        0 & 1 & 0 \\
        -s_{13}e^{i\delta} &0 & c_{13}
    \end{pmatrix}
    \begin{pmatrix}
        c_{12} & s_{12} & 0\\
        -s_{12} & c_{12} & 0\\
        0 & 0 & 1
    \end{pmatrix}
    \begin{pmatrix}
        1 & 0 & 0\\
        0 & e^{i\frac{\alpha_{21}}{2}} & 0\\
        0 & 0 &  e^{i\frac{\alpha_{31}}{2}}
    \end{pmatrix}\,.
\end{aligned}
\end{equation}
Here, $c_{ij} \equiv \cos\theta_{ij}$, $s_{ij} \equiv \sin\theta_{ij}$, with $\theta_{ij}$ being the mixing angles of the active neutrinos, and $\delta$ is the Dirac phase, whereas $\alpha_{21}, \alpha_{31}$ are the Majorana phases. The matrix $R$ is complex orthogonal~\cite{Granelli:2020pim}
\begin{equation}
    R=\begin{pmatrix}
        1 & 0 & 0 \\
        0 & c_{\omega_{1}} & s_{\omega_{1}} \\
        0 &- s_{\omega_{1}} & c_{\omega_{1}} 
    \end{pmatrix}
    \begin{pmatrix}
        c_{\omega_{2}} & 0 & s_{\omega_{2}} \\
        0 & 1 & 0\\
        -s_{\omega_{2}} & 0 & c_{\omega_{2}} 
    \end{pmatrix}\\
    \begin{pmatrix}
        c_{\omega_{3}} & s_{\omega_{3}} & 0\\
        -s_{\omega_{3}} & c_{\omega_{3}} & 0\\
        0 & 0 & 1
    \end{pmatrix}\,,
\end{equation}
where $c_{w_{i}} \equiv \cos\omega_{i}$, $s_{w_{i}} \equiv \sin\omega_{i}$, and $\omega_{i} = x_i + i y_i$ are complex angles.

For simplicity we assume that one of the RHN decouples, with a negligible contribution to light neutrino masses. In this scenario, the angles $w_{1,3} = \pi/2$ are no longer free while $w_2$ remains undetermined~\cite{Granelli:2020ysj},

For the light neutrinos, we assume a normally ordered scenario with the lightest neutrino massless. We use the oscillation parameters~\cite{Esteban:2020cvm} 
\begin{gather}
\label{eq:osc-parameters}
    \theta_{12} = 33.44^{\circ} \quad
    \theta_{13} = 8.57^{\circ}, \quad
    \theta_{23} = 49.20^{\circ}, \nonumber\\
    m_{\nu_2} = \sqrt{\Delta m^2_\text{sol}} 
              = 8.6 \times 10^{-3}~\text{eV}, \quad
    m_{\nu_3} = \sqrt{\Delta m^2_\text{atm}} 
              = 5.0 \times 10^{-2}~\text{eV}.
\end{gather}

Once $\lambda_D$ is determined, since $N$ only decays to SM particles via the mixing to the active neutrinos, the decay width of the $N \rightarrow \ell_\ell H, \bar{\ell}_\ell H^{*}$ can be expressed as $\Gamma_{N_j} = \frac{M_{N_j}}{8\pi} \sum_\ell [\lambda_D^\dagger]_{j\ell}[\lambda_D]_{\ell j}$. 

\subsection{$CP$ Asymmetry}
In order to generate the BAU, a model needs to satisfy the three conditions of Sakharov~\cite{Sakharov:1967dj}. First, lepton number violation processes are introduced via the decays of the Majorana RHNs. Second, the Yukawa couplings of the RHNs are $C$ and $CP$ violating. Third, RHN decays fall out of thermal equilibrium, generating a lepton asymmetry. While sphaleron processes are still active, the lepton asymmetry will be transferred to a baryon asymmetry, which can be expressed as~\cite{Buchmuller:2004nz}
\begin{equation}
\label{eq:etaB}
    \eta_B^f \simeq 10^{-2}\times \sum_{\ell} N_{\Delta_\ell}(z = z_\text{sph})
    \simeq 10^{-2}\times\sum_{i,\ell} \varepsilon_{i\ell} \, \kappa_{i\ell}(z_{\text{sph}}),
\end{equation}
where $z = M_{N_1}/T$, with $M_{N_1}$ being the mass of the lightest RHN. $z_{\text{sph}} = M_{N_1} / T_\text{sph}$ with $T_\text{sph} \approx 130$~GeV is the sphaleron freeze-out temperature~\cite{Burnier:2005hp}. $N_{\Delta_{\ell}}$ is the number density in a comoving volume of $\Delta_{\ell} = B/3-L_{\ell}$ for $\ell =e, \mu, \tau$. $\varepsilon_{i\ell}$ is the magnitude of the $CP$ asymmetry,
\begin{equation}
    \epsilon_{i\ell} = \frac{\Gamma_{N_i\to\ell_\ell H}-\Gamma_{N_i\to\bar\ell_\ell H^*}}{\Gamma_{N_i\to\ell_\ell H}+\Gamma_{N_i\to\bar\ell_\ell H^*}}.
\end{equation}
In Eq.~(\ref{eq:etaB}), $\kappa_{i\ell}$ are the thermal efficiencies which are determined by solving the relevant Boltzmann equations shown below in Eq.~(\ref{eq:bolt}).

If two of the RHNs have mass degeneracy, such that their mass difference is of a similar order as their decay widths, $\Delta M_{N} = |M_{N_2}-M_{N_1}| \approx \Gamma_N$, then the $CP$ asymmetry is resonantly enhanced~\cite{DeSimone:2007edo}~\footnote{However, this expression does not consider the divergence in the one-loop approximation in the exactly degenerate limit. Out-of-equilibrium QFT methods are required to describe this more accurately~\cite{Klaric:2021cpi}.}~\footnote{Thermal corrections are not considered as well, which has small contribution to the final asymmetry in the case of strong washout, i.e. $K_i \gtrsim 1$~\cite{Giudice:2003jh}.
More accurate description can be found in Ref.~\cite{Frossard:2012pc, Giudice:2003jh, Hambye:2016sby}.
In the following discussions, our selected benchmark satisfy the strong washout regime.}%
\begin{align} 
\label{eq:CP}
-\epsilon_{i \ell}&=\sum_{j\neq i} 
        \frac{\operatorname{Im}\left([\lambda_D^\dagger]_{i\ell}
                [\lambda_D]_{\ell j}[\lambda_D^\dagger \lambda_D]_{ij}\right) 
        + \frac{M_{N_i}}{M_{N_j}} \operatorname{Im}\left([\lambda_D^\dagger]_{i
                \ell} [\lambda_D]_{\ell j}[\lambda_D^\dagger \lambda_D]_{ji}\right)}{[\lambda_D^\dagger\lambda_D]_{ii}[\lambda_D^\dagger
        \lambda_D]_{jj}} 
        \left(f_{ij}^\text{mix} + f_{ij}^\text{osc}\right),
\end{align}
with
\begin{align} 
    f_{ij}^{\operatorname{mix}} =
    \frac{(M_{N_i}^2 - M_{N_j}^2) M_{N_i}\Gamma_{N_j}}
    {(M_{N_i}^2 - M_{N_j}^2)^{2} + M_{N_i}^2\Gamma_{N_j}^2}, 
\end{align}
and
\begin{align} 
    f_{ij}^{\mathrm{osc}} =
    \frac{(M_{N_i}^2 - M_{N_j}^2) M_{N_i}\Gamma_{N_{j}}}
    {(M_{N_i}^2 - M_{N_j}^2)^2 + (M_{N_i}
    \Gamma_{N_i} + M_{N_j} \Gamma_{N_j})^2
    \frac{\operatorname{det}\left[\operatorname{Re}(\lambda_D^\dagger
    \lambda_D)\right]}{[\lambda_D^\dagger \lambda_D]_{ii}[\lambda_D^\dagger \lambda_D]_{jj}}}\,.  
\end{align} 
Here, $i,j=1,2$, and we assume the third RHN so heavy that it decouples. 

In the given scenario with two RHNs and adopting normally ordered light neutrino masses, the $CP$ asymmetry $\varepsilon_{i\ell}$ depends on six free parameters:
\begin{itemize}
    \item The mass of the lightest RHN, $M_N \equiv M_{N_1} \approx M_{N_2}$; 
    \item the mass difference between the two degenerate RHNs, $\Delta M_N \equiv |M_{N_2} -M_{N_1}|$,
    \item the real and imaginary parts, $x_2$ and $y_2$, of the complex angle $\omega_2$; 
    \item the Dirac $CP$ phase and the difference between the two Majorana $CP$ angles, $\alpha_{23} \equiv \alpha_{21}-\alpha_{31}$.
\end{itemize}
The other parameters are fixed by neutrino oscillation data, where we take the best fitted values as in Ref.~\cite{Esteban:2020cvm}.

\begin{figure}[t!]
    \centering
    \includegraphics[width=0.495\textwidth]{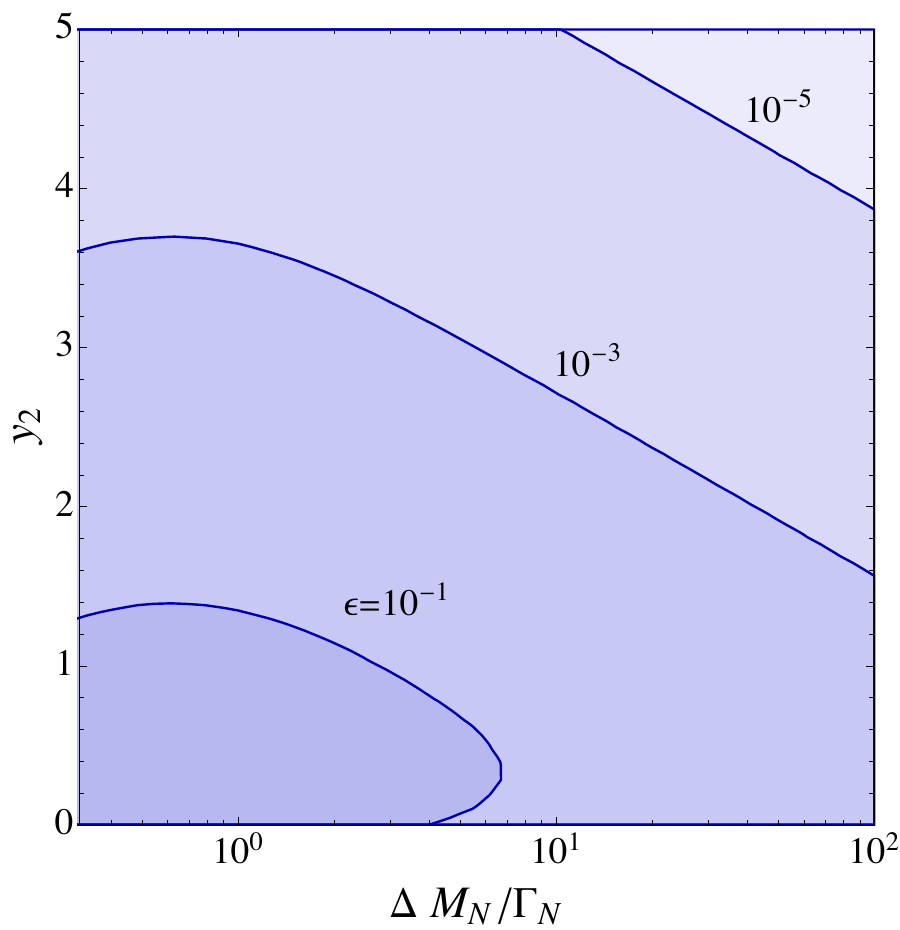}
    \includegraphics[width=0.485\textwidth]{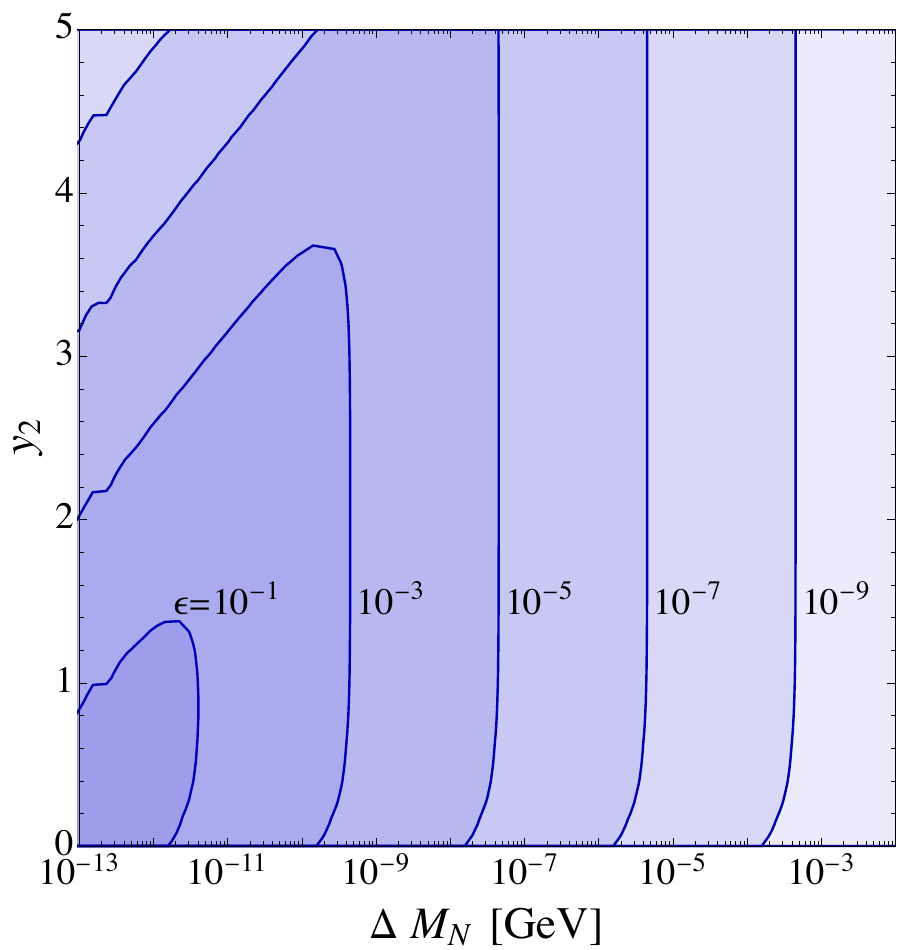}
    \caption{$CP$ asymmetry $\epsilon_{1\mu}$ as a function of $y_2$, as well as $\Delta M_N / \Gamma_N$~(left) and $\Delta M_N$~(right). We fix the other parameters as $\delta = 3\pi/2$, $\alpha_{23} = \pi$, $x_2 = \pi/4$ and $M_N = 100$~GeV.}
\label{fig:CP}
\end{figure}
In order to illustrate the parameter dependence of the $CP$ asymmetry, Fig.~\ref{fig:CP} shows it as a function of $\Delta M_N / \Gamma_N$~(left) and $\Delta M_N$~(right) as well as $y_2$. The other four parameters are fixed to be $\delta = 3\pi/2$, $\alpha_{23} = \pi$, $x_2 = \pi/4$ and $M_N = 100$~GeV. Since $\Gamma_N$ increases as $y_2$ gets larger, so in Fig.~\ref{fig:CP}~(left), $\Delta M_N$ increases with larger $y_2$ with fixed $\Delta M_N / \Gamma_N$. That is why the contours in Fig.~\ref{fig:CP}~(right) are steeper than in Fig.~\ref{fig:CP}~(left). We show $\epsilon_{1\mu}$ as an example, and the two generations of the RHNs have similar $CP$ asymmetry. Therefore, we observe that $\epsilon_{i\ell}$ gets larger when the mass difference $\Delta M_N$ approaches the decay width, $\Delta M_N \sim \Gamma_N \approx 10^{-13}$~GeV. As the imaginary part $y_2$ of the complex angle in the $R$ matrix gets larger, $\epsilon$ drops sharply. By naive approximation of Eq.~(\ref{eq:etaB}), in order to get $\eta_B^f = \eta_B^\text{ob} \simeq 6.1 \times 10^{-10}$~\cite{ParticleDataGroup:2016lqr}, since $\kappa \lesssim 1$, we require a $CP$ asymmetry $\epsilon_{i\ell} \gtrsim 10^{-8}$, which will require $\Delta M_N \lesssim 10^{-5}$~GeV from Fig.~\ref{fig:CP}. 

\subsection{Boltzmann Equations and Lepton Number Washout}

In order to proceed, we need to calculate the thermal efficiencies $\kappa_{i\ell}$ to obtain the final BAU. Comparing to the standard case of resonant leptogenesis, the Boltzmann equations contain an additional term due to $NN \to Z' \to f\bar{f}$ scattering~\cite{Heeck:2016oda}~\footnote{We focus on scenarios where $M_N \ll M_Z'$ and $M_\Phi$, hence other scattering processes including $NN \to Z' Z', Z' \Phi, \Phi\Phi$ are sub-leading~\cite{Heeck:2016oda}.,
\begin{align}
\label{eq:bolt}
    \frac{{\rm d}N_{N_i}}{{\rm d}z} &= 
    -(D(K_i)+S_{h/A}
    )(N_{N_i}-N_{N_i}^{\rm eq}) 
    - 2\,S_{Z'}/N_{N_i}^{\rm eq}(N_{N_i}^2-(N_{N_i}^{\rm eq})^2)\; , \nonumber\\
    \frac{{\rm d}N_{\Delta_{\ell}}}{{\rm d}z} &= \sum_i \varepsilon_{i\ell} \, D(K_i)\, (N_{N_i}-N_{N_i}^{\rm eq})-\sum_i W^{\rm 0}(K_{i\ell}) N_{\Delta_{\ell}}\; ,
\end{align} 
where $N_{N_i~(\Delta_{\ell})}$ are the relevant number densities in the comoving volume~($i=1,2$). $N_{N_i}^{\rm eq}$ is the equilibrium number density of $N_i$, }
\begin{align}
    N_{N_i}^{\rm eq}(z)=\frac{1}{2} z^2 \mathcal{K}_2(z),\,
\end{align} 
where $\mathcal{K}_i(z)$ is the modified Bessel function of the $i$-th type. 

After taking into account thermal corrections ~\cite{Hambye:2016sby}, $D(K_i)$ is the decay term
\begin{align}
\label{eq:dd}
    D(K_i) &= \theta(M_N-M_H-M_L) K_i z \mathcal{K}_1(z) \frac{1}{\mathcal{K}_2(z)} \lambda^{1/2}[1,a_H,a_L](1-a_H+a_L) \nonumber\\
    &+ \theta(M_H-M_N-M_L)K_i z \mathcal{K}_1 \left(\frac{M_H}{M_N}z \right)\frac{1}{\mathcal{K}_2(z)} \frac{M_H^2}{M_N^2} \lambda^{1/2}[a_H,1,a_L] \frac{(a_H-a_L-1)}{a_H^{3/2}}, 
\end{align}
where $\theta(x)$ is the Heaviside step function, $a_X = (M_X(T)/M_N)^2$ and $\lambda[a,b,c] = (a-b-c)^2-4bc$. For the thermal corrections on the Higgs mass, we take $M_H^2(T) \approx M_H^2(v(T)) + (\frac{y_t^2}{4} + \frac{\lambda}{2})T^2$ \cite{Giudice:2003jh}, with $v^2(T) = (1-T^2/T_c^2)\theta(T_c-T)v^2$, $T_c = 160$~GeV and $v=246$~GeV~\cite{Kajantie:1995kf}. Since we are only interested in $T > T_{\text{sph}}$, we have $M_H(T) = 0.632 T$ and $M_\ell(T) =  0.296 T$~\cite{Granelli:2020ysj}, while the thermal corrections to RHN masses are negligibly small~\cite{Giudice:2003jh}. The above expression reflects the two kinematic regions $M_N(T) > M_H(T) + M_L(T)$ and $M_H(T) > M_N(T) + M_L(T)$.
The quantity $K_i$ represents the decay parameter
\begin{align}
\label{decpari}
    K_i &= [\lambda_D^{\dagger}\lambda_D]_{ii} v_{\text{EW}}^2 / (M_{N} m_{\star}) 
    = \tilde{m}_i/m_{\star},
\end{align}
where $\tilde{m}_i = [\lambda_D^\dagger\lambda_D]_{ii} v_{\text{EW}}^2 / M_N$ is the effective neutrino mass, and $m_\star = \frac{16\pi^{5/2\sqrt{g_\star}}}{3\sqrt{5}} \frac{v_{EW}^2}{M_\text{pl}} \simeq 1.08\times 10^{-3}$~eV, is the equilibrium neutrino mass~\cite{Buchmuller:2004nz}. Here, $g_* = 106.75$ is the number of relativistic degrees of freedom, and $M_\text{pl} = 1.22\times 10^{19}$~GeV is the Planck scale. 

The $\Delta L = 1$ scattering processes are mediated by both $H$ and $A$,
\begin{align}
    S_{h/A} = 2S_{h/A,s} + 4S_{h/A,t},
\end{align}
with detailed expressions given in \cite{Giudice:2003jh} and an appropriate approximation in \cite{Granelli:2020ysj}. Furthermore, $W^0$ is the total washout including contributions from scattering and inverse decays~\cite{Buchmuller:2004nz}
\begin{align}
    W^{\rm 0}(K_{i \ell}) = W^{\rm ID}(K_{i\ell}) + W^{\Delta L=1},
\end{align} 
with
\begin{align}
    W^{\Delta L=1} = W^{h,s} + 2W^{h,t},
\end{align} 
and expressions thereof given in \cite{Buchmuller:2004nz}. $W^{\rm ID}(K_{i\ell})$ is the washout term from the inverse decay of the $i$-th RHNs to lepton $\ell$,
\begin{align}
    W^{\rm ID}(K_{i \ell},z)=\frac{1}{4}K_{i\ell}\mathcal{K}_1(z)z^3,
\end{align} 
with
\begin{align}
\label{decpar}
    K_{i\ell} = \frac{\widetilde{\Gamma}_{\rm D} (N_i\to L_{\ell} H + \bar{L}_{\ell}
    H^\dagger)}{H(z=1)} 
    = \frac{[\lambda_D^\dagger\lambda_D]_{\ell i} v_\text{EW}^2}{M_N m_\star},
\end{align}
with the Hubble rate defined through $H^2 = (8\pi/3 M_{\rm pl}^2)\rho$, $\rho = \frac{\pi^2}{30}g_{*}T^4 = \frac{\pi^2}{30}g_{*}(M_N / z)^4$.

Likewise, $S_{Z^\prime}$ is the scattering term mediated by $Z^\prime$,
\begin{align}
    S_{Z^\prime} \equiv \gamma_{Z'}/(H N_N^{\rm eq} z),
\end{align} 
with $\gamma_{Z^\prime}$ is the reaction rate of the scattering process via $Z^\prime$~\cite{Blanchet:2009bu},
\begin{align}
    \gamma_{Z^\prime} = \frac{M_N}{64\pi^4z}\int_{s_{\rm min}}^\infty ds\,
    \hat\sigma_{Z'}(s)\sqrt{s}K_1\left(\frac{\sqrt{s}}{M_N}z\right).
\end{align} 
Here, the reduced cross section is
\begin{align}
    \hat{\sigma}_{Z'}(s) = 
    \frac{13 g_{B-L}^4}{6\pi}
    \frac{\sqrt{s(s-4 M_N^2)^3}}{(s-M_{Z'}^2)^2 + M_{Z'}^2\Gamma_{Z'}^2},
\end{align} 
with the $Z^\prime$ decay width
\begin{align}
    \Gamma_{Z'} = 
    \frac{g_{B-L}^2}{24\pi}M_{Z'}
    \left(13 + 3(1-4M_N^2/M_{Z^\prime}^2)^{3/2}\right),
\end{align} 

When the scattering processes are efficient, the densities of the RHNs closely follow that in thermal equilibrium, hence we can take $N_{N_i} \approx N_{N_i}^\text{eq}$, thus Eq.~(\ref{eq:bolt}) can be solved to yield the efficiency factor~\cite{Blanchet:2009bu},
\begin{align}
\label{eq:efficiency}
    \kappa_{i\ell} (z,z_{\rm in}) \approx 
    \int^z_{z_{\rm in}}&{\rm d}z'
    \frac{{\rm d}N_{N_i}^{\rm eq}}{{\rm d}z'}
    \frac{D(K_i,z')}{D(K_i,z')+S_{h/A}(z')+4S_{Z'}(z')} \nonumber\\
    &\times\exp\left[-\int^{z}_{z'} {\rm d}z'' \sum_i W^{\rm 0}(K_{i\ell},z'') \right],
\end{align}
with $W^{\rm 0} = W^{\rm ID}(D+S_{H/A})/D$. We take $z_{\rm in} = {\rm min}[10^{-2},0.1 T_\text{sph}]$ at the start of leptogenesis, since for $z \ll 1$, the RH neutrinos are too close to thermal equilibrium and the contribution to efficiency is small.

To numerically calculate the final BAU, including scattering effects, we use the Python package {\tt ULYSSES}~\cite{Granelli:2020pim} to determine the $CP$ asymmetry $\epsilon_{i \ell}$ and the decay parameter $K$. We then use Eq.~(\ref{eq:efficiency}) to calculate the efficiency. Finally, we use Eq.~(\ref{eq:etaB}) to derive the final BAU. In summary, the introduction of the scattering $NN \to Z^\prime \to f \bar{f}$ pushes the decays of the RHNs back to thermal equilibrium, hence this decreases the efficiency and subsequently, the BAU. The final BAU is hence also controlled by the cross section of the scattering, which depends on $M_{Z^\prime}$ and $g_{B-L}$.

\begin{figure}[t!]
    \centering
    \includegraphics[width=0.60\textwidth]{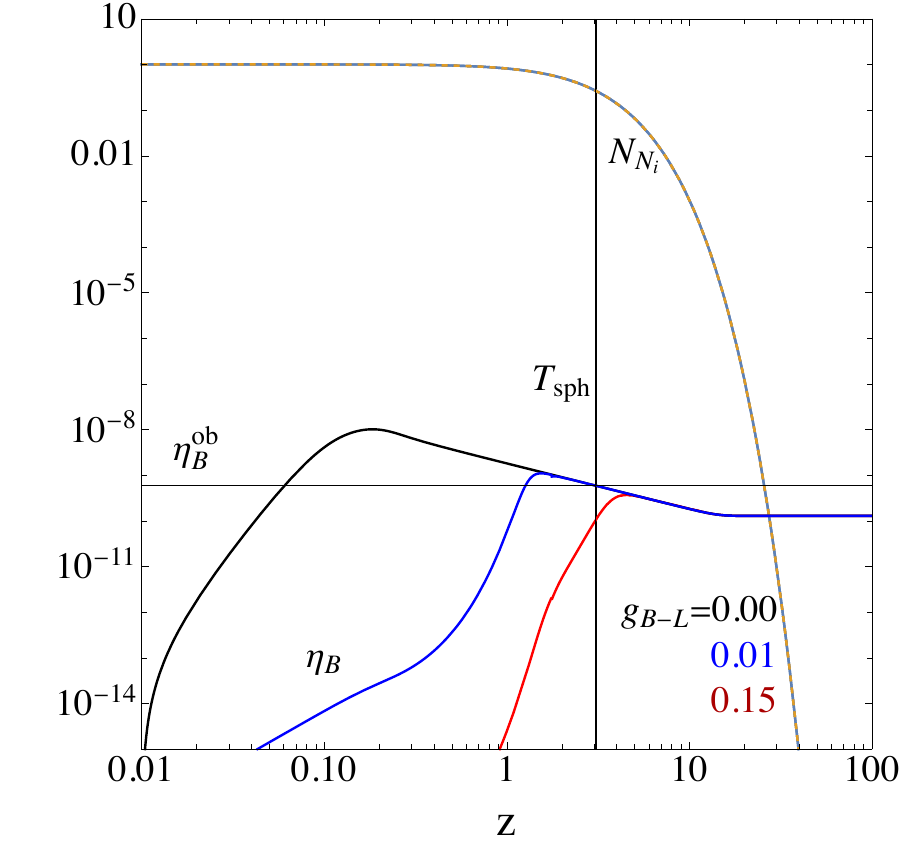}
    \caption{RHN number density $N_{N_1}$ and the BAU $\eta_B$ as a function of $z = m_N / T$, for different values of the $B-L$ gauge coupling $g_{B-L} = 0$, $0.01$, $0.15$. The other parameters are $M_{Z'} = 5 \times 10^3$~GeV, $M_N = 400$~GeV, $\Delta M_N = 0.05 \Gamma_N \approx 10^{-10}$~GeV, $y_2 = 3.4$, $x_2 = \pi/4, \delta = 3\pi/2$, and $\alpha_{23} = \pi$. The final BAU, $\eta_B^f$ is equal to $\eta_B$ at $T = T_\text{sph}$, and is compared to the observed BAU by the intersection of the sphaleron freeze-out temperature, $T_\text{sph} \approx 130$~GeV and $\eta_B^\text{ob} = 6.1\times 10^{-10}$. }
\label{fig:evo}
\end{figure}
As an example, in Fig.~\ref{fig:evo}, we show the density of $N_{N_1}$ and the BAU, $\eta_B(z) \approx 10^{-2} \times \sum_{\ell}N_{\Delta_{\ell}}(z)$ as a function of $z$, for four different benchmark points with $M_{Z^\prime} = 5 \times 10^3$~GeV, and $g_{B-L} = 0$, 0.01, and $0.15$. These benchmark points are chosen according to the current experimental limits as shown in Fig.~\ref{fig:limit}. The other parameters are chosen as $M_N = 400$~GeV, $\Delta M_N = 0.05 \Gamma_N \approx 10^{-10}$~GeV, $y_2 = 3.4$, and the default parameters for $x_2, \delta, \alpha_{23}$ are to be shown in Eq.~\ref{eq:param}. Thermal corrections and $\Delta L=$1 scattering terms are not considered here. Later, we are going to discuss the LLP signatures at colliders for these benchmark points as well. As shown in the figure, larger $g_{B-L}$ turns out to decrease the final BAU obtained at $z=z_{\text{sph}} = M_N / T_\text{sph}$ as expected. For $g_{B-L} = 0.15$, the final BAU is over $10$ times smaller than in the decoupled case with $g_{B-L} = 0$.

In total, we have eight free parameters in resonant leptogenesis within the $B-L$ model. Six of them affect the $CP$ asymmetry $\epsilon_{i\ell}$, and two the scattering which results in different thermal efficiencies, $\kappa$. The parameters can be mapped to the flavour-summed active-sterile mixing~\cite{Klaric:2021cpi} 
\begin{align}
\label{eq:v2}
    |V|^2 \equiv \Sigma_{i=1,2, \ell=e,\mu,\tau} |V_{\ell N_i}|^2 
    = \frac{\Sigma_i m_{\nu_i}}{M_{N}} \cosh(2y_2) \approx 2 \Sigma_{\ell=e,\mu,\tau} |V_{\ell N_{1/2}}|^2.
\end{align}
The decay length of lightest RHN is then $\Sigma_{\ell=e,\mu,\tau} |V_{\ell N_1}|^2 \equiv |V_1|^2$.

The active-sterile mixing depends on the Yukawa couplings. Larger Yukawa couplings will lead to larger washout rates. Hence, successful leptogenesis limits large Yukawa couplings as well as large $|V|^2$. The largest viable $|V|^2$ is obtained when the Dirac $CP$ phase $\delta$, the Majorana $CP$ phase $\ell$, and $x_2$ satisfy the relations~\cite{Klaric:2020phc, Klaric:2021cpi,Drewes:2021nqr}
\begin{align}
\label{eq:param}
    \delta = n \pi/2, \quad 
    \alpha_{23} = m\pi, \quad 
    x_2 = l\pi/4
    \quad\text{with}\quad m,n,l \in \mathbb{Z}\,.
\end{align}

In our analysis we are interested in the largest allowed parameter space for successful leptogenesis and light neutrino mass generation in the $B-L$ model. The connection to LLP searches will be discussed in the next section. We fix and vary the model parameters as
\begin{gather}
\label{eq:parameters}
    M_N = 1 - 1000~\text{GeV}, \quad 
    \Delta M_N  = 10^{-17} - 10^{-4}~\text{GeV}, \nonumber\\
    M_{Z'} = 5~\text{TeV}, \quad 
    g_{B-L} = 0.025, 0.05, 0.15, \\
    x_2 = \pi/4, \quad 
    y_2 = 0 - 5, \quad
    \delta = 3\pi/2, \quad
    \alpha_{23} = \pi. \nonumber
\end{gather}
Together with the oscillation parameters in Eq.~\eqref{eq:osc-parameters}, the PMNS matrix is then fixed to
\begin{equation}
    U = \begin{pmatrix}
        0.825         & 0.545i        & 0.149i \\
        -0.360+0.094i & 0.062+0.545i  & 0.749  \\
        0.417+0.081i  & 0.054 -0.632i & 0.646
    \end{pmatrix},
\end{equation}
and the active-sterile mixing $|V_{\ell N_i}|^2 = (v_\text{EW}[y_D]_{\ell i} / M_{N_i})^2$ is given by
\begin{align}
    &|V_{\ell N_i}|^2 \approx 10^{-13}
     \times\frac{100~\text{GeV}}{M_N} \nonumber\\
    &\times 
    \begin{pmatrix}
       |0.59 e^{-1.57i}\cosh y_2+0.12 \sinh y_2|^2  & |0.12 e^{1.57 i} \cosh y_2+0.59 \sinh y_2|^2 & 0 \\
        |1.20e^{0.3i}\cosh y_2+1.28 e^{1.29i} \sinh y_2|^2 & 
        |1.28e^{-0.28i}\cosh y_2+1.20 e^{-1.27i} \sinh y_2|^2 & 0 \\
        |1.07e^{-0.4i}\cosh y_2-1.14 e^{-1.20i} \sinh y_2|^2 & 
        |1.14e^{0.37i}\cosh y_2-1.07 e^{1.17i} \sinh y_2|^2 & 0
    \end{pmatrix}.
\end{align}

In the next section, we are going to discuss the collider phenomenology especially the LLP signatures, and its connection to the seesaw and leptogenesis.

\section{Long-lived Particle Searches}
\label{sec:llp}

For RHNs with GeV to TeV scale masses, colliders are the most promising environment for detection. While prompt final states have been mainly searched for due to easier triggering and reconstruction, displaced final states originating from LLPs have attracted increasing interest both experimentally and theoretically \cite{Alimena:2019zri}. 

Despite the heavier RHN masses considered in our case, RHNs are most likely long-lived for two reasons: (a) The $CP$ asymmetry $\epsilon_{i \ell}$ is required to be sufficiently large to make it potentially detectable by searching for same-sign dilepton signals. This requires small $y_2$ hence small active-sterile mixing as shown in Fig.~\ref{fig:CP}. (b) In the canonical seesaw, as $|V_{\ell N}|^2 \simeq m_\nu / M_N$, the active-sterile mixing is also small leading to long RHN decay lengths.
 
For these values of the active-sterile mixing and $M_N \sim \mathcal{O}(1-10)$~GeV, their decay length is~\cite{Atre:2009rg, Deppisch:2018eth}
\begin{align}
    \label{lengthapproxi-onshell-light}
    L_N \approx 5 \times 10^4 ~\text{m}
    \times\frac{10^{-12}}{|V|^2}
    \times\left(\frac{10~\text{GeV}}{M_N}\right)^5
    \times\frac{\beta\gamma}{10^3} \approx \frac{8 \times 10^3 ~\text{m}}{\cosh(2y_2)}\times \left(\frac{10~\text{GeV}}{M_N}\right)^4\times\frac{\beta\gamma}{10^3},
\end{align}
Likewise, for larger masses $M_N \sim \mathcal{O}(10^2)$~GeV and heavier, the decay length is~\cite{Atre:2009rg, Liu:2022kid} 
\begin{align}
    \label{lengthapproxi-onshell}
    L_N \approx 2~\text{cm}
    \times\frac{10^{-12}}{|V|^2}
    \times\left(\frac{100~\text{GeV}}{M_N}\right)^3
    \times\frac{\beta\gamma}{100} \approx \frac{3.2~\text{cm}}{\cosh(2y_2)}\times \left(\frac{100~\text{GeV}}{M_N}\right)^2\times\frac{\beta\gamma}{100},
\end{align}
where we have taken account that $|V|^2 \approx 2 |V_{1/2}|^2$.
The above relation is normalized for a RHN boost $\beta\gamma = 10^3~(100)$. This is the approximate average factor in $Z'$ production, $pp \to Z' \to NN$, for $m_{Z'} = 5$~TeV at the 100~TeV FCC-hh \cite{Liu:2022kid}, for $M_N \sim 10~(100)$ GeV.

One of the most striking signatures of Majorana RHNs are same-sign dileptons. Such processes also allow for the $CP$ asymmetry $\epsilon_{i\ell}$ to be measured experimentally \cite{Liu:2021akf}. The HL-LHC can potentially probe $\epsilon_{i\ell}\sim 0.1$ by measuring an asymmetry in dilepton production. Regarded as a very optimistic outlook, we here discuss the future 100~TeV FCC-hh to search for RHNs \cite{Liu:2022kid}. With 30 ab$^{-1}$ of integrated luminosity and a higher collision energy, we expect the FCC-hh to be able to probe even smaller $CP$ asymmetries. As shown in Fig.~\ref{fig:CP}, $\epsilon_{i\ell}\lesssim 0.1$ is only obtained for $y_2\lesssim 1.5$ which corresponds to $|V|^2 \lesssim 10^{-12}$, for $M_N\sim \mathcal{O}(100)$~GeV.

\begin{figure}[t!]
    \centering
    \includegraphics[width=0.49\textwidth]{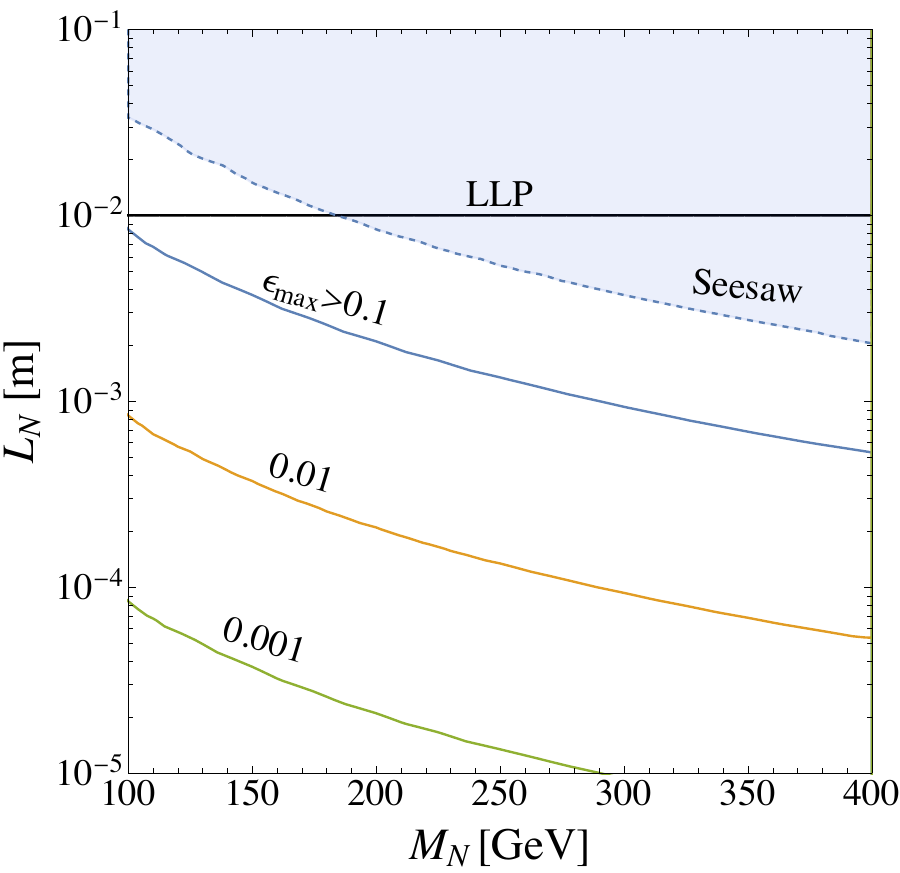}
    \caption{Maximal $CP$ asymmetry $\epsilon_\text{max}$ in muon flavour (left) and the maximal BAU $\eta_B^f$ without $Z'$ washout (right), as a function of $M_N$ and laboratory RHN decay length assuming a boost $\beta\gamma = 100$. The canonical seesaw floor is indicated by the shaded region. The horizontal line labelled 'LLP' delineates the approximate minimal decay length for detectable LLPs.}
\label{fig:ld}
\end{figure}
In Fig.~\ref{fig:ld}, using the benchmark parameters as in Eq.~\eqref{eq:param} and focusing on 100 GeV $< M_N < 400$ GeV~\footnote{This mass range is selected so the parameter space of seesaw floor and large $CP$ asymmetry is close to the boundary of the decay length where RHNs can be regarded as a LLP.}, we show the maximal $CP$ asymmetry $\epsilon_\text{max}$ in muon flavour, as a function of $M_N$ and the decay length $L_N$ in the laboratory frame, where we assume a boost factor $\beta\gamma = 100$ at the FCC-hh. As the $CP$ asymmetry is a function of $\Delta M_N$, the maximal $CP$ asymmetry is obtained when $\Delta M_N \sim \Gamma_N$. The canonical seesaw floor is at $|V|^2 = \Sigma_i m_{\nu_i} / M_N$ with $\Sigma_i m_{\nu_i} \approx 0.06$~eV. The horizontal line labelled 'LLP' indicates $L_N > 0.01$~m where the RHNs can be approximately regarded as LLPs \cite{Liu:2022kid}. We can see that a detectable $CP$ asymmetry $\epsilon_{i\ell}\sim \mathcal{O}(0.1)$, require a laboratory decay length $L_N\gtrsim 10^{-3}$~m. Likewise, the canonical seesaw floor occurs where $L_N\gtrsim 10^{-2}$~m, and the RHNs can be detected in LLP signatures at the FCC-hh. We do not expect sufficient $CP$ asymmetry in the prompt region where the laboratory decay length is small, especially for light RHNs. Hence, we can expect LLP searches at the FCC-hh to test leptogenesis and light neutrino mass generation simultaneously.

LLP searches at the FCC-hh using the process $pp \to Z' \to NN$ have been considered in Ref.~\cite{Liu:2022kid}. The LLPs are searched for using the tracker, calorimeter and muon systems of the main detectors of the FCC-hh. RHNs dominantly decay into lepton plus jets.  Once they decay within $L_N \sim 0.01-10$~m, they are regarded as detected. No background is assumed, since the RHNs are long-lived enough. As a very optimistic outlook, a perfect reconstruction efficiency is assumed. Such a search is mainly sensitive to muon flavour, $|V_{\mu N}|^2$. In order to simplify our discussion, we take the benchmark parameters as in Eq.~\eqref{eq:param}. In Fig.~\ref{fig:flavora23}~(top left), we find that in this case, $|V_{\mu N}|^2$ is about 10 times larger than the other components. Hence, we take $|V|^2 \approx 2|V_1|^2 \approx 2|V_2|^2 \approx 2|V_{\mu N}|^2$. This allows us to relate the LLP searches depending on $|V_{\mu N}|^2$ for the two essentially degenerate RHNs with the overall mixing strength $|V|^2$ relevant for leptogenesis.

\begin{figure}[t!]
    \centering
    \includegraphics[width=0.99\textwidth]{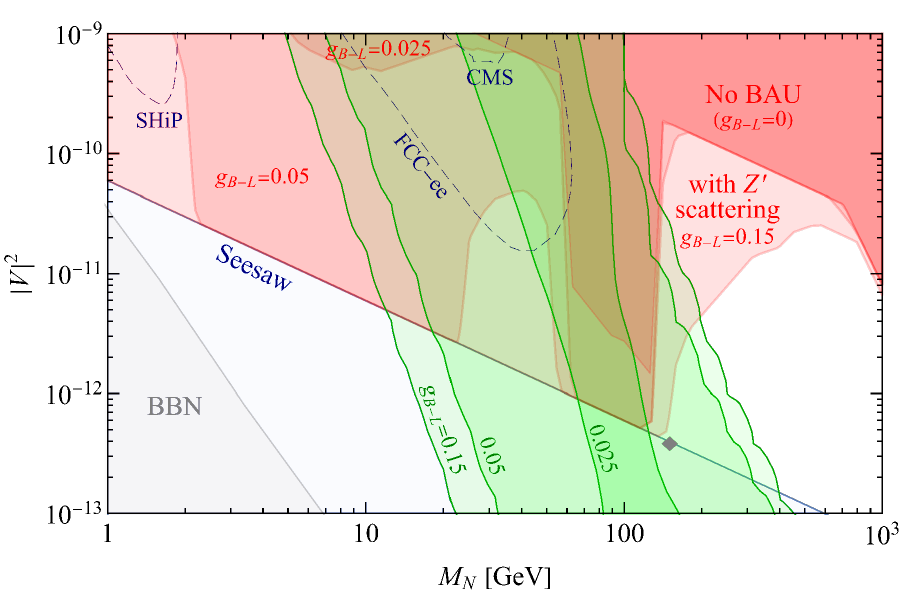}
    \caption{Projected sensitivity at 95\% confidence level on RHN LLP searches for the process $pp \to Z' \to NN$ at the FCC-hh~\cite{Liu:2022kid} (diagonal green band). The $Z'$ mass is $M_{Z'} = 5$~TeV and the $B-L$ gauge coupling assumes three different values $g_{B-L} = 0.025$, $0.05$ and $0.15$ as indicated. The LLP search is based on detection of muons and we have $|V|^2 \approx 2|V_{1,2}|^2 \approx 2|V_{\mu N}|^2$ when fixing the Majorana $CP$ phase $\alpha_{23} = \pi$.
    No sufficient baryon asymmetry is generated in the upper right corner labelled 'No BAU' ($g_{B-L} = 0$) whereas the other three red shaded regions indicate where the observed BAU cannot be achieved due to $Z'$ washout ($g_{B-L} = 0.025$, $0.05$ and $0.15$). Below the line labelled 'Seesaw', active neutrino masses cannot be generated at a sufficient level, $|V|^2 < 0.06~\text{eV}/M_N$. For comparison, projected sensitivities for sterile neutrino searches at SHiP~\cite{SHiP:2018xqw}, CMS~\cite{Drewes:2019fou} and FCC-ee~\cite{Blondel:2014bra, Blondel:2022qqo} are shown as well, and the region labelled 'BBN' is disfavoured due to the impact of sterile neutrinos on big bang nucleosynthesis. The diamond indicates a benchmark scenario discussed below.}
\label{fig:str}
\end{figure}
Hence, we correlate the parameter space for successful leptogenesis and observable LLP searches in Fig.~\ref{fig:str}. The BAU is calculated using Eq.~\eqref{eq:etaB}, with the efficiency in Eq.~\eqref{eq:efficiency} and the $CP$ asymmetry in Eq.~\eqref{eq:CP}. The region labelled \textit{'No BAU'} has a final asymmetry $\eta_B^f < \eta_B^\text{ob}$ and it is thus excluded due to strong washout from the Yukawa couplings, even without the presence of the additional $Z'$ washout in the $B-L$ model\footnote{This description is not exact for large couplings \cite{Klaric:2020phc, Klaric:2021cpi, Drewes:2021nqr}, where the use of quantum kinetic equations including the RHN coherence terms is required. In the region of interest, such corrections are expected to be less relevant due to the additional $Z'$ scattering. The contribution of the coherence terms will be smaller since the RHNs are not required to be so degenerate. This is especially the case if the maximal $CP$ asymmetry is large.}. The region labelled \textit{'Seesaw'} corresponds to the parameter space where the seesaw mechanism fails to generate the active neutrino masses $\Sigma_i m_{\nu_i} \approx 0.06$~eV. For comparison, we show the projected sensitivities of sterile neutrino searches at SHiP~\cite{SHiP:2018xqw}, CMS~\cite{Drewes:2019fou} and FCC-ee~\cite{Blondel:2014bra, Blondel:2022qqo}, with data taken from \url{http://www.sterile-neutrino.org/} \cite{Bolton:2019pcu, Bolton:2022pyf}. It is important to emphasize that these are based on the induced RHN interactions due to SM currents and the active-sterile mixing, i.e., without the additional $Z'$ interactions. Thermal corrections become important for temperatures near and below the EW phase transition, i.e., $M_N \lesssim 100$~GeV. This has been taken into account in Eq.~\ref{eq:dd}. As shown in the figure, for $M_N \sim M_H$, the thermal corrections to the Higgs and lepton masses are critical, such that the decays $H \to L + N$ and $N \to L + H$ are both suppressed. Hence, the viable parameter space for leptogenesis shrinks, even excluded totally for $g_{B-L} \gtrsim 0.025$.
Nevertheless, an asymmetry generated from sterile neutrino oscillations has not been considered, which is also important for such low masses. Our results are thus less reliable for $M_N \lesssim 100$~GeV. We still plot it as the excluded region approximately extrapolated for lighter RHNs, as we are mainly interested in the parameter space near the canonical seesaw floor and to demonstrate that the additional $Z'$ scattering will rule out most of the parameter space anyway.

The red regions are excluded due to strong $Z'$ washout for $M_{Z'} = 5$~TeV and three different values of $g_{B-L} = 0.025$, $0.05$, $0.15$. The inclusion of the scattering processes $NN \to Z' \to f\bar{f}$ decrease the thermal efficiency, therefore additional parameter space is ruled out by requiring the BAU to meet the observation. The scattering is sizeable for lighter RHNs and, as expected, the excluded region becomes larger for increasing $g_{B-L}$.

The green band can be probed using LLP searches at the FCC-hh in the process $pp\to Z' \to NN$, again for $M_{Z'} = 5$~TeV and the three values $g_{B-L} = 0.025$, $0.05$, $0.15$. It is thus the desired region for phenomenological observation. The band width increases with larger $g_{B-L}$ due to a more abundant production of RHNs. By comparing the contours and the BAU excluded regions corresponding to the same $g_{B-L}$, we can determine if neutrino mass generation and leptogenesis can be probed simultaneously, for a certain $g_{B-L}$ value. For the parameter space outside the BAU excluded region, but inside the LLP contours, leptogenesis can be probed. For the parameter space inside the BAU excluded region and the LLP contours, leptogenesis can be potentially falsified. Below the seesaw floor but inside the green LLP contours, neutrino mass generation in the $B-L$ model can be falsified. Specifically, for $g_{B-L} = 0.025$, LLP searches can potentially falsify neutrino mass generation, but can only falsify leptogenesis nearing $M_N \sim 100$ GeV. On the other hand, for $g_{B-L} = 0.05$, $Z'$ washout is effective for $M_N \approx 20 - 60$~GeV. The LLP contours and the BAU excluded region now intersect over a broad range. If the $Z^\prime$  and RHN were to be detected in LLP searches, resonant leptogenesis in our scenario could be falsified, assuming the relevant masses and couplings can be determined to a sufficient precision. For $g_{B-L} = 0.15$, i.e., near the current upper limit, cf., Fig.~\ref{fig:limit}, the effects of $Z'$ scattering are so strong that resonant leptogenesis is not possible for $M_N \lesssim 110$~GeV.

\begin{figure}[t!]
    \centering
    \includegraphics[width=0.45\textwidth]{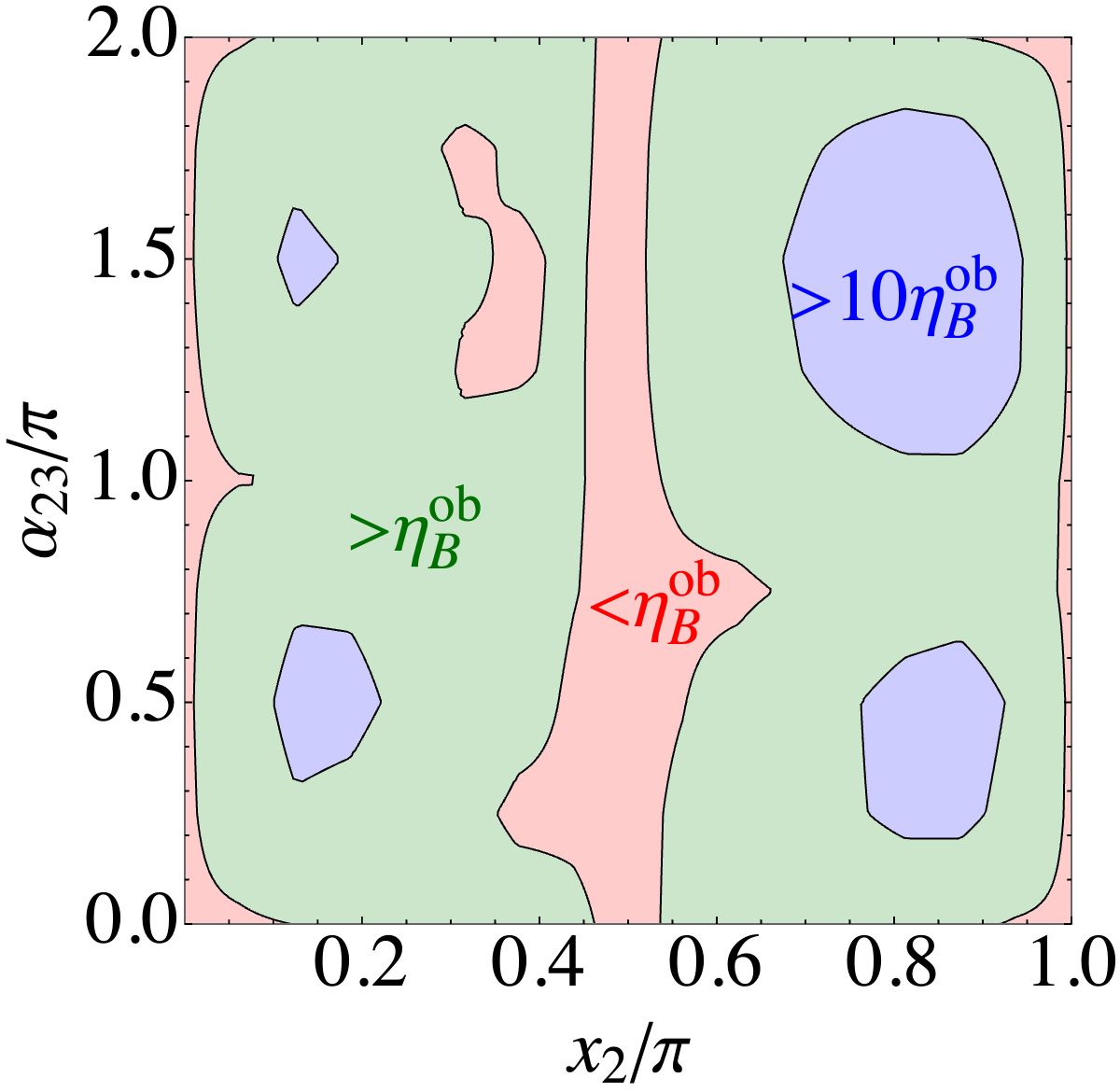}  \includegraphics[width=0.45\textwidth]{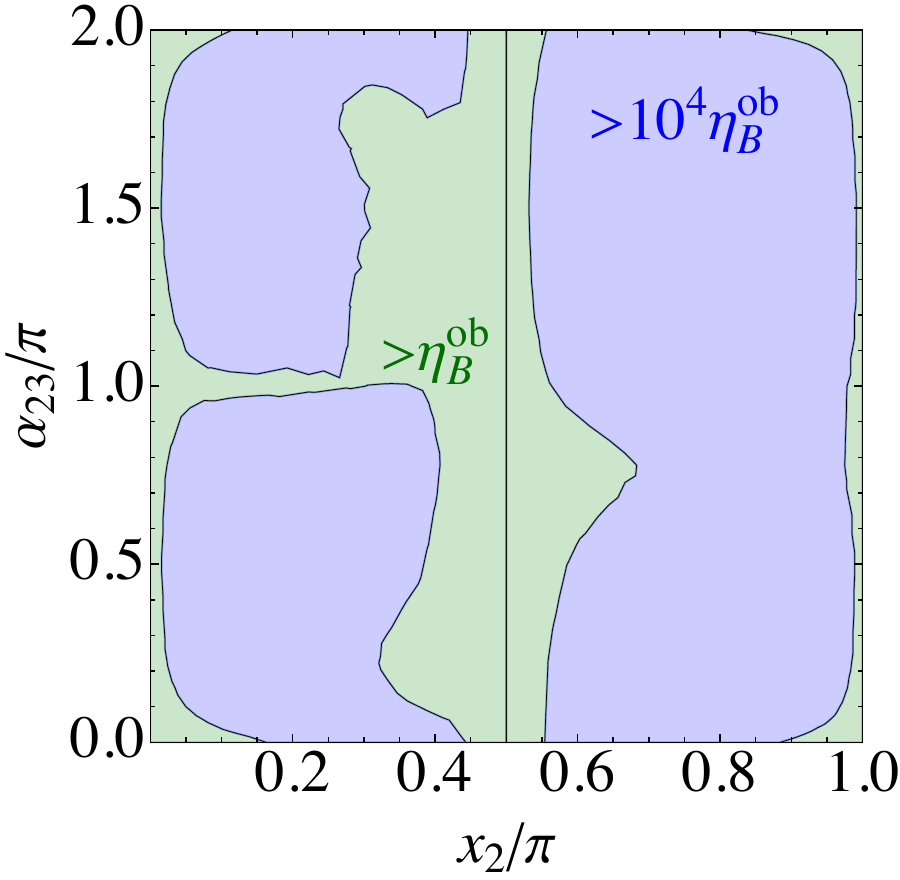}
    \caption{Maximal BAU achievable as a function of the real part $x_2$ of the $R$-matrix angle and the Majorana phase $\alpha_{23}$. The other parameters are fixed to $M_N = 150$~GeV, $|V|^2 \approx 4\times 10^{-13}$, and the Dirac CP phase $\delta = 3\pi/2$ and mass difference $\Delta M_N$ are chosen such that the $CP$ asymmetry is maximal. The left plot is under the presence of $Z'$ washout with $M_{Z'} = 5$~TeV and $g_{B-L} = 0.15$ whereas the right plot is for standard resonant leptogenesis in the sterile neutrino extended SM.}
\label{fig:ob}
\end{figure}
In a potential discovery, if the $Z'$ and RHN were to be detected in direct resonance and LLP searches, further details on the viability of leptogenesis can be elucidated. In Fig.~\ref{fig:ob}~(left) we show the largest BAU achievable as a function of real part $x_2$ of the $R$-matrix angle and the Majorana phase $\alpha_{23}$. We assume that the other parameters are determined as $M_{Z'} = 5$~TeV, $g_{B-L} = 0.15$ (e.g., in a direct resonance search) as well as $M_N = 150$~GeV, $|V|^2 \approx 4\times 10^{-13}$ (left). This scenario, indicated by the diamond in Fig.~\ref{fig:str}, enables successful generation of light neutrino masses, being exactly on the minimal active-sterile mixing required. Since it is hard or impossible to independently measure the $CP$ phases, $x_2$ and the mass difference $\Delta M_N$, we treat them as remaining free parameters. Specifically, we choose $\Delta M_N$ and the Dirac $CP$ phase to maximize the BAU. This compares with Fig.~\ref{fig:ob}~(right), showing the maximal BAU achievable without $Z'$ scattering, i.e., for standard resonant leptogenesis in the sterile-neutrino-extended-SM, where all of the parameter space is viable. Hence, although $Z'$ washout decreases the BAU by more than three orders of magnitude, leptogenesis can still be successful in a wide parameter space in Fig.~\ref{fig:ob}~(left), and requiring it further constrains the parameter space. Observing a signal in scenarios with strong washout can falsify leptogenesis as a viable mechanism of BAU generation. For example, for $M_N = 30$~GeV, $|V|^2 = 2\times 10^{-12}$ and the other parameter kept the same, leptogenesis cannot achieve the observed BAU, as can be seen in Fig.~\ref{fig:str}.

\section{Conclusion}
\label{sec:con}
The observed baryon asymmetry of the Universe and non-zero neutrino masses both point towards the existence of physics beyond the Standard Model. Extending the SM, the $B-L$ model is one of the simplest ultraviolet-complete scenarios to explain the the baryon asymmetry via resonant leptogenesis and neutrino masses via the type-I seesaw mechanism. Both are triggered by the presence of heavy sterile neutrinos, or RHNs. In this context, the $B-L$ model has the additional benefit that the heavy Majorana masses of the RHNs are generated by the spontaneous breaking of the $B-L$ gauge symmetry, and thus connected to the mass scale of the exotic $Z'$ gauge boson. If this breaking occurs around the $TeV$ scale, EW-scale RHN masses are naturally generated. While suggestive of resonant leptogenesis, such RHNs are difficult to probe via the tiny active-sterile mixing $|V|^2 \gtrsim 0.06~\text{eV}/M_N \approx 10^{-12}$ at $M_N = 100$~GeV, but in the $B-L$ model, they can be produced more abundantly at colliders via the $Z'$, e.g., $pp \to Z' \to NN$. This allows probing very small active-sterile mixing strength via LLP searches, even below the seesaw floor. On the other hand, the $Z'$ interactions also lead to increased washout of lepton and baryon number in leptogenesis. This suggests an interesting interplay between leptogenesis and LLP searches in the $B-L$ model.

In this work, we discuss the potential of using LLP searches to test both resonant leptogenesis and neutrino mass generation in the $B-L$ model. Both mechanisms favor for long-lived RHNs, since in leptogenesis the $CP$ asymmetry is required to be sufficiently large to be potentially detectable, which requires small active-sterile mixing. Regarding the LLP searches, we consider the projected sensitivity at a future 100~TeV FCC-hh \cite{Liu:2022kid} to explore the most optimistic prospects. We have considered three benchmark scenarios, with $M_{Z'} = 5$~TeV, and $g_{B-L} =0.025$, $0.05$ and $0.15$. In all cases, LLP searches for $pp \to Z' \to NN$ at the FCC-hh are expected to have a powerful sensitivity on the active-sterile mixing $|V|^2$, reaching the parameter space well below the seesaw floor. Such sensitivity of the seesaw floor can be expected for even smaller couplings, as long as $g_{B-L} \gtrsim 0.002$. For the case where the coupling is strong enough, $g_{B-L} =$ 0.15, the LLP searches are sensitive to the parameter space within 10 $ \lesssim M_N \lesssim 120$~GeV, where the BAU and neutrino masses both can not be sufficiently generated by leptogenesis and the seesaw mechanism, making the scenario falsifiable. Probing the scenarios in more detail, anticipating a positive signal of $CP$ violation in neutrino oscillations, neutrinoless double beta decay and an exotic signal at the FCC-hh can thus enable illuminating why there is matter in the universe.
 
\acknowledgments
The authors thank Ye-ling Zhou for useful discussions. WL is supported by National Natural Science Foundation of China (Grant No.12205153). FFD acknowledges support from the Science and Technology Facilities Council, part of U.K. Research and Innovation, Grant No. ST/T000880/1 and ST/X000613/1.

\appendix
\section{Dependence of $|V_{\ell N}|^2$ on model parameters}
\label{sec:flavour}
Experimentally, individual RHNs are searched for via their couplings with certain flavours of leptons. Limits and sensitivities are put on the individual $|V_{\ell N_i}|^2$, instead of their sum as relevant for leptogenesis. Therefore, it is necessary to discuss the magnitude of different flavour components of the $|V|^2$ matrix in order to correlate experimental searches with leptogenesis. The active-sterile mixing for each flavour and RHN can be expressed as
\begin{align}
    |V_{\ell N_i}|^2 \approx 
    |(v_{\text{EW}} (\lambda_{D})_{\ell i}/M_{N_i})|^2, .
\label{eq:vllambda}
\end{align}
with $\ell = e, \mu, \tau$ and $i = 1,2$ (there is no summation over $i$). As the two RHNs are highly degenerate, we have $|V_{\ell N}|^2 \equiv |V_{\ell N_1}|^2 = |V_{\ell N_2}|^2$.  Below, we plot the active-sterile mixing  as a function of the model parameters as listed in Eq.~\eqref{eq:parameters}. 

\begin{figure}[t!]
    \centering
    \includegraphics[width=0.49\textwidth]{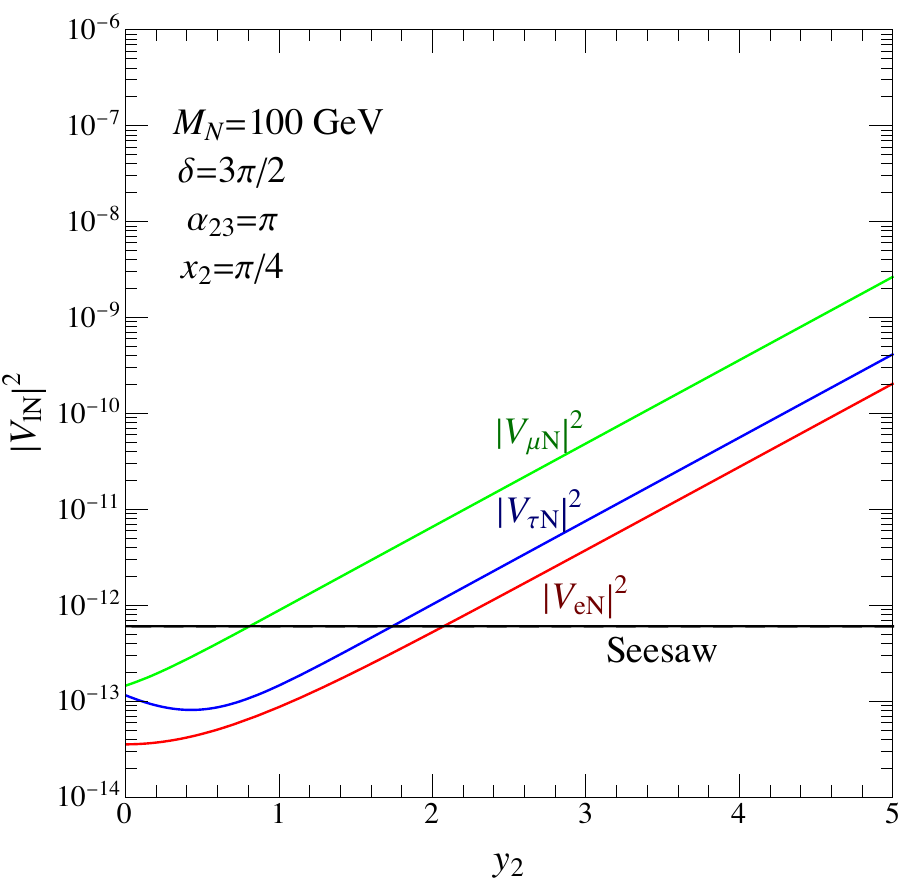}
    \includegraphics[width=0.49\textwidth]{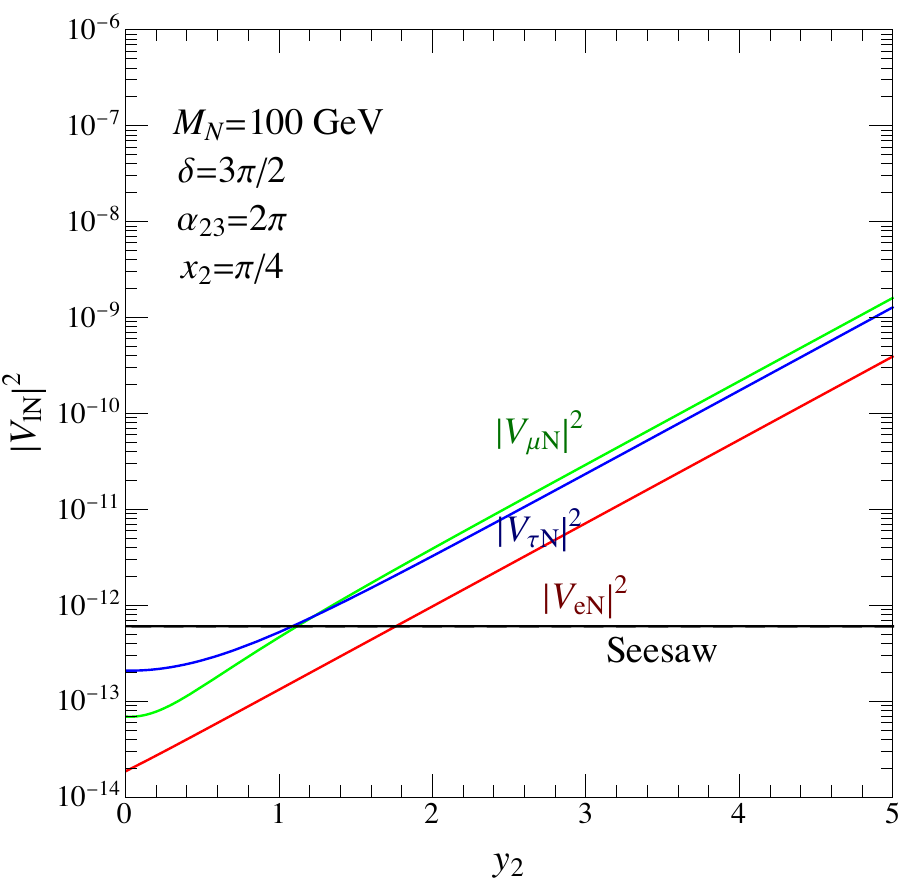}\\
    \includegraphics[width=0.49\textwidth]{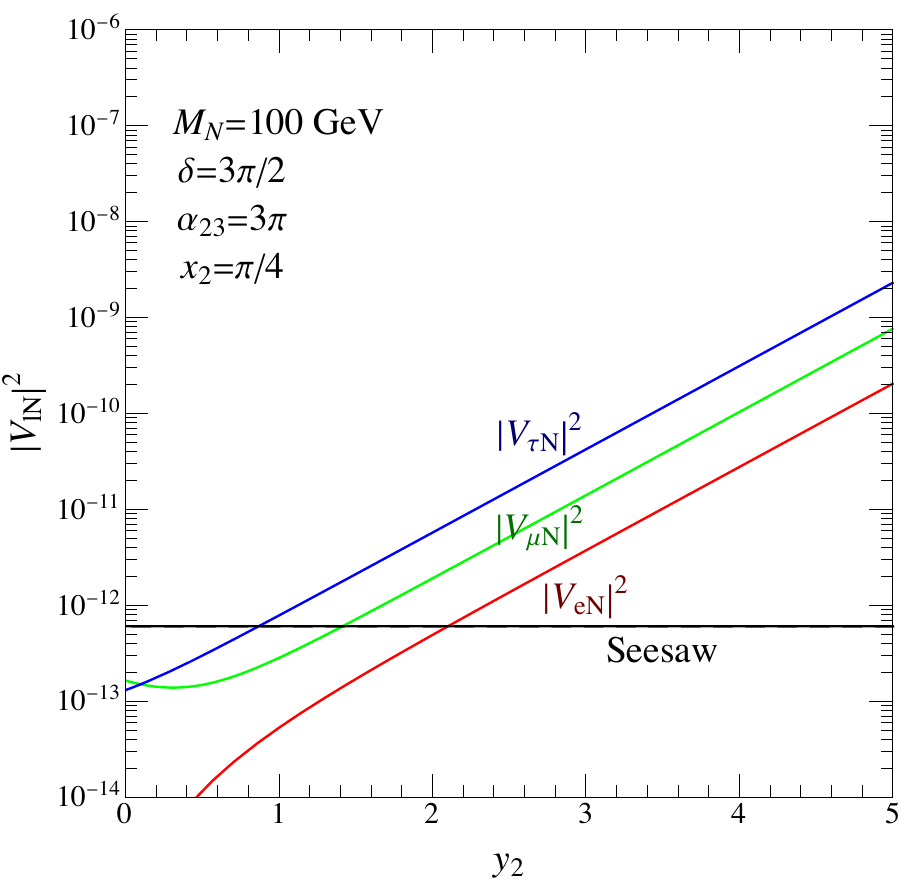}
    \includegraphics[width=0.49\textwidth]{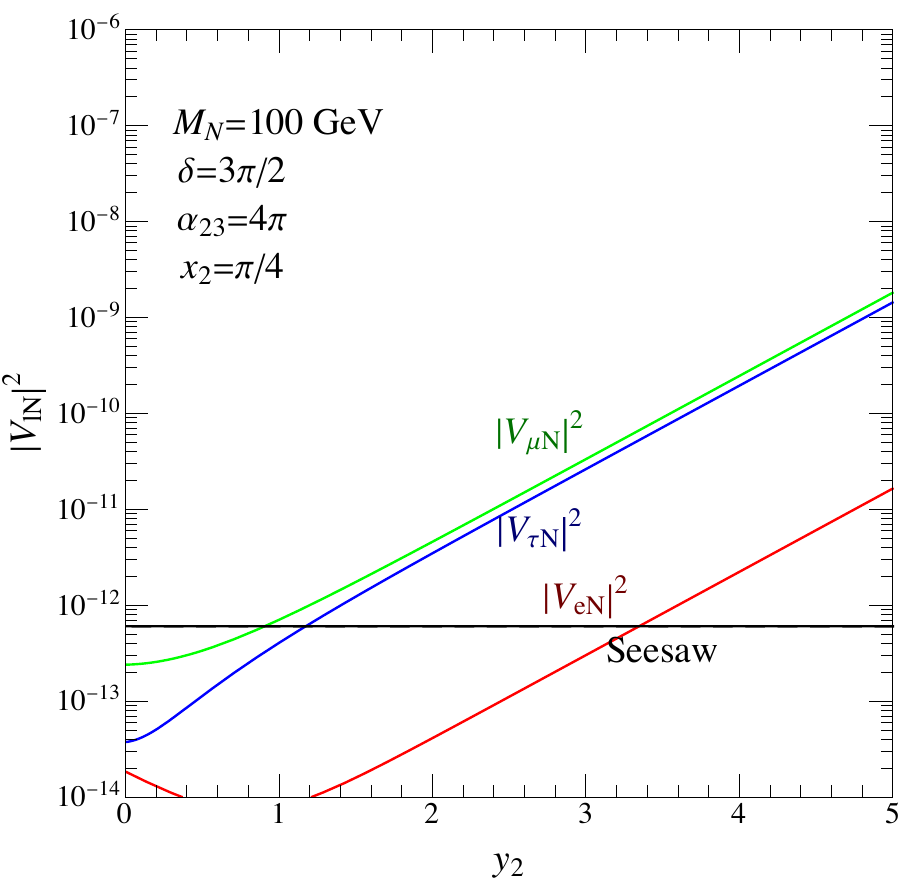}
    \caption{Active-sterile mixing strengths $|V_{\ell N}|^2$~($\ell = e, \mu, \tau$) as a function of $y_2$, for $\alpha_{23} = \pi$ (top left), $2\pi$ (top right), $3\pi$ (bottom left) and $4\pi$ (bottom right) with the other parameters as indicated. The value in the canonical seesaw case, $|V_{\ell N}|^2 = m_\nu / M_N$, is indicated by the horizontal line.}
\label{fig:flavora23}
\end{figure}
In Fig.~\ref{fig:flavora23}, we show $|V_{\ell N}|^2$ as a function of $y_2$, for a different values of the Majorana $CP$ phase $\alpha_{23}$. The other free parameters are fixed as $M_N = 100$~GeV, $\delta = 3\pi/2$ and $x_2 = \pi/4$, while the mass difference $\Delta M_N$ is not relevant. The canonical seesaw floor is indicated for comparison. The 'seesaw' line is put to reflect the canonical seesaw floor, $|V_{\ell N}|^2 \approx \sum_i m_{\nu_i} / M_N$. In most cases, $|V_{\ell N}|^2$ increases with $y_2$, except certain cases when $y_2 \lesssim 1$. Comparing the four panels with different $\alpha_{23}$ changing from $\pi$ to $4\pi$, the dominant flavour of $|V_{\ell N}|^2$ is controlled by $\alpha_{23}$: For $\alpha_{23} = \pi$, $|V_{\mu N}|^2$ is the dominant flavour component, which is about one magnitude larger with sufficiently large $y_2$. However, when $\alpha_{23} = 2\pi$ and 4$\pi$, $|V_{\mu N}|^2$ and $|V_{\tau N}|^2$ are comparable, and $|V_{\tau N}|^2$ becomes dominant when $\alpha_{23} = 3\pi$. In all cases, $|V_{e N}|^2$ is smallest. 

\begin{figure}[t!]
    \centering
    \includegraphics[width=0.49\textwidth]{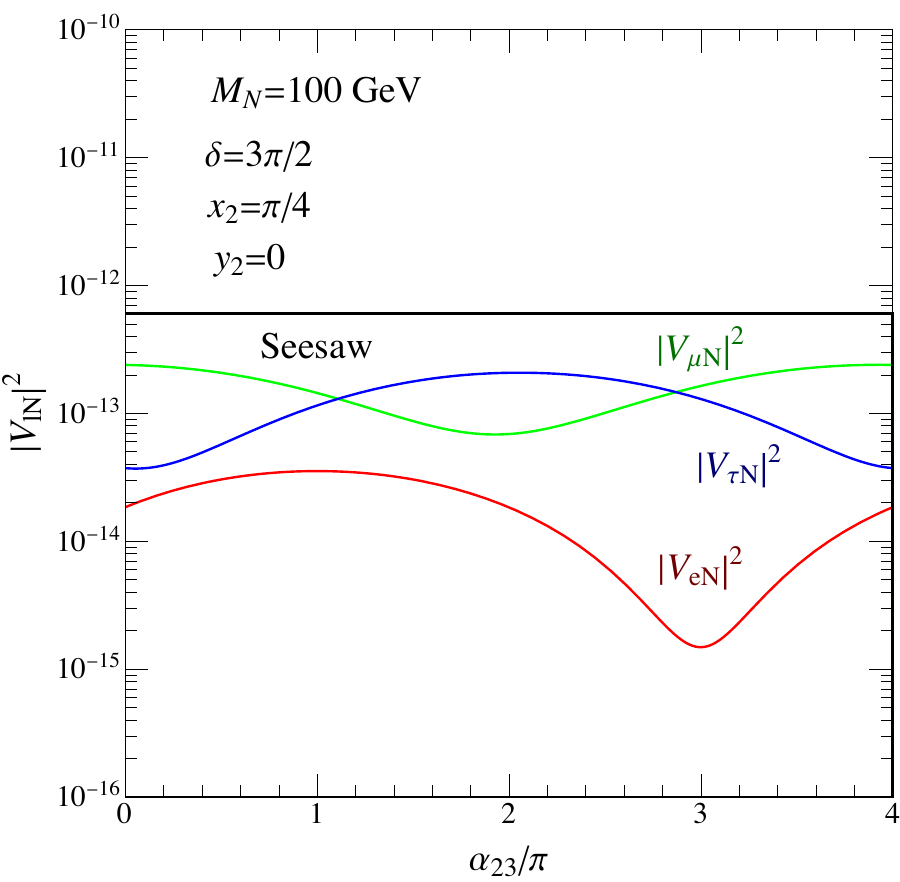}
    \includegraphics[width=0.49\textwidth]{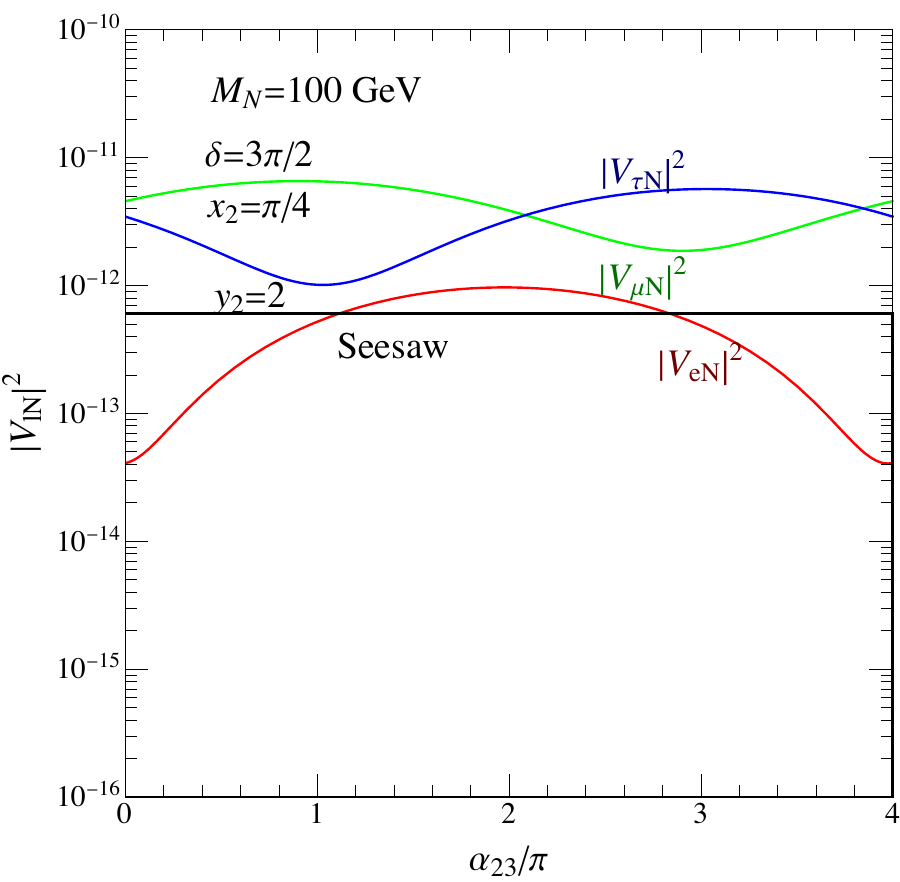}
    \caption{As Fig.~\ref{fig:flavora23}, but showing dependence on $\alpha_{23}$, for $y_2 = 0$ (left) and 2 (right).}
\label{fig:flavora23y}
\end{figure}
\begin{figure}[t!]
    \centering
    \includegraphics[width=0.49\textwidth]{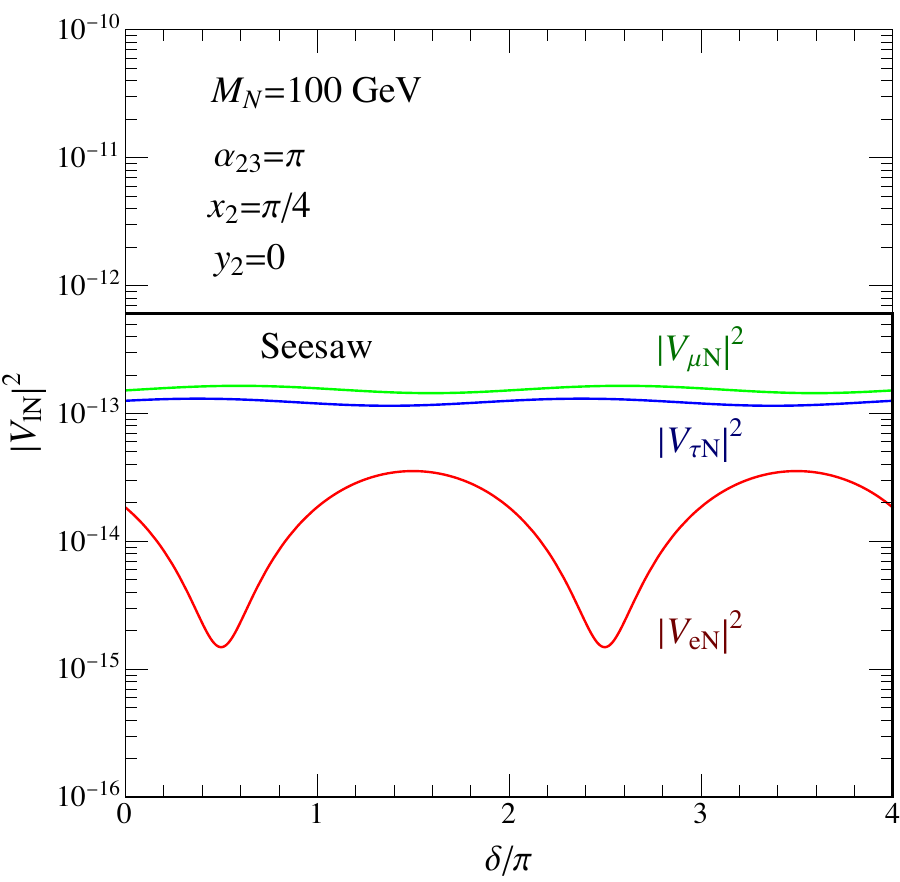}
    \includegraphics[width=0.49\textwidth]{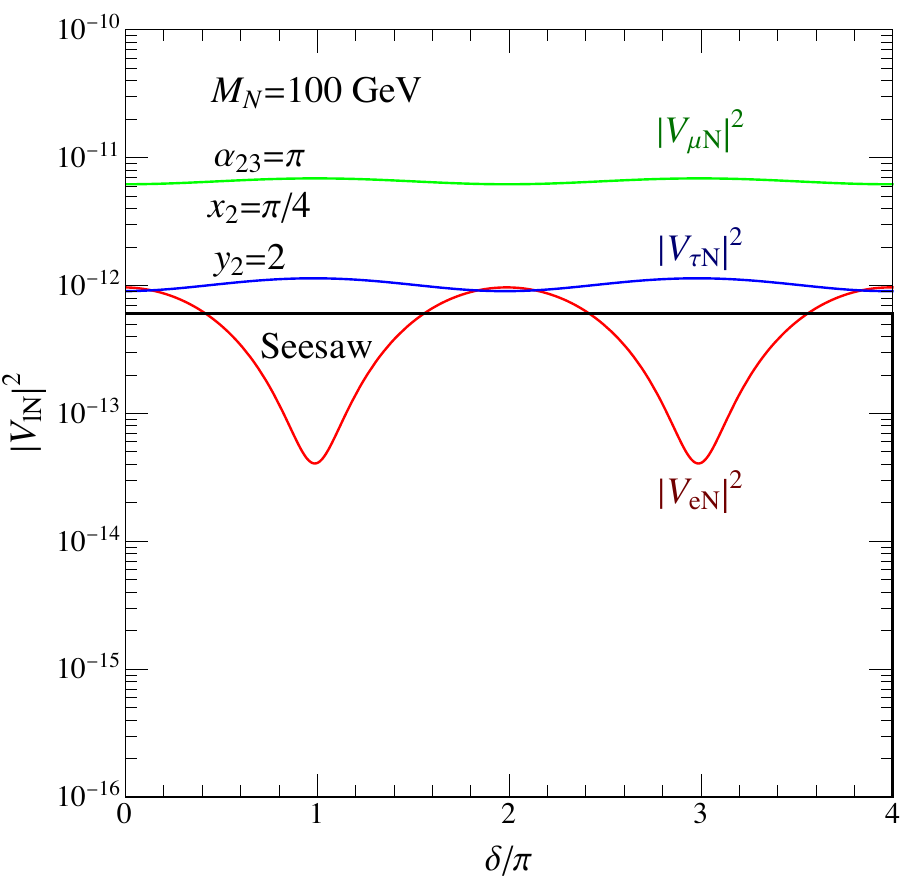}
    \caption{As Fig.~\ref{fig:flavora23}, but showing dependence on $\delta$, for $y_2 = 0$ (left) and 2 (right).}
\label{fig:flavordelta}
\end{figure}
\begin{figure}[t!]
    \centering
    \includegraphics[width=0.49\textwidth]{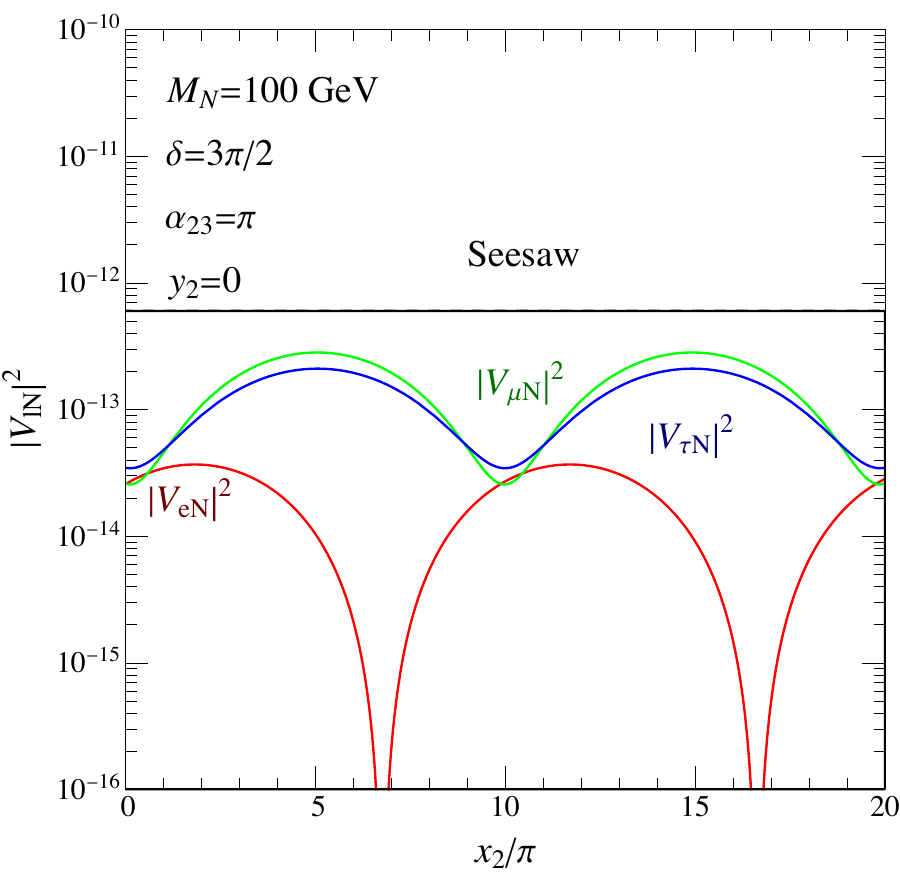}
    \includegraphics[width=0.49\textwidth]{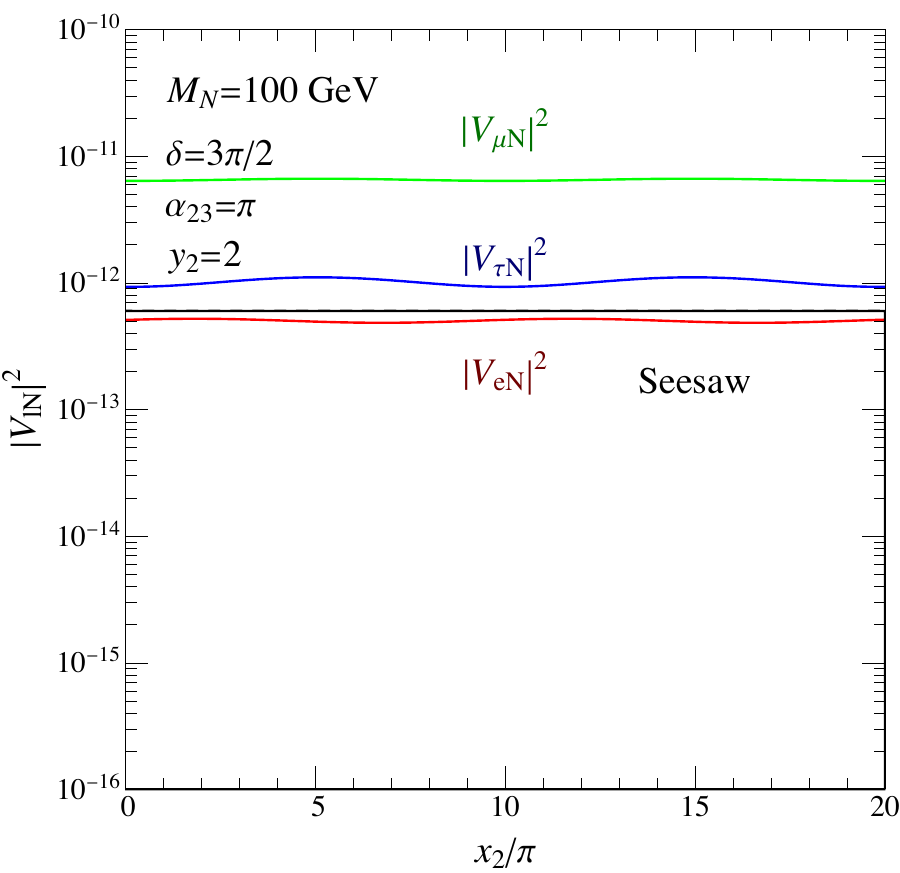}
    \caption{As Fig.~\ref{fig:flavora23}, but showing dependence on $x_2$, for $y_2 = 0$ (left) and 2 (right).}
\label{fig:flavorx2}
\end{figure}

Focusing on the dependence of $|V_{\ell N}|^2$ on $\alpha_{23}$, we demonstrate this in Fig.~\ref{fig:flavora23y} for $y_2 = 0$ and 2. Likewise, Figs.~\ref{fig:flavordelta} and \ref{fig:flavorx2} show the dependence on $\delta$ and $x_2$, respectively. We observe that $|V_{\mu N}|^2$ and $|V_{\tau N}|^2$ vary least with $\delta$ and $x_2$. However, $|V_{e N}|^2$ is smallest in most cases. Hence, the dominant flavour component $|V_{\ell N}|^2$ does not change appreciably with $\delta$ and $x_2$.
Thus, $|V_{\ell N}|^2$ mainly depends on $y_2$ and $\alpha_{23}$, and the dominant flavour component is $|V_{\mu N}|^2$ or $|V_{\tau N}|^2$, depending on $\alpha_{23}$.

\bibliographystyle{JHEP}
\bibliography{submit}
\end{document}